\documentclass[pra,aps,superscriptaddress,showpacs,preprint]{revtex4}%
\usepackage{amsmath}
\usepackage{amssymb}
\usepackage{amsfonts}
\usepackage{bm}
\usepackage{amssymb}
\usepackage[dvips]{graphicx}
\usepackage{dcolumn}
\usepackage{txfonts}
\usepackage{hyperref}
\usepackage{makeidx}
\usepackage{color}
\usepackage{mathtools}
\usepackage{mathrsfs}
\usepackage{subfigure}
\usepackage{appendix}

\begin{document}

\title{Synthetic cooling translational mode of an optically trapped nanoparticle through librational mode}
\author{Ke-Wen Xiao}
\affiliation{Beijing Computational Science Research Center, Beijing 100193, China}

\author{Anda Xiong}
\affiliation{School of Physics and Astronomy, University of Birmingham, Birmingham, UK}

\author{Nan Zhao}
\affiliation{Beijing Computational Science Research Center, Beijing 100193, China}

\author{Zhang-qi Yin}\email{yinzhangqi@mail.tsinghua.edu.cn}
\affiliation{Center for Quantum Information, Institute for Interdisciplinary Information Sciences, Tsinghua University, Beijing 100084, China}

\begin{abstract}
We systematically investigate the multi-stability behaviour and cooling of both librational and translational modes of an optically levitated nonspherical nanoparticle. By expanding the trapping potential to the fourth order of both the translational and librational freedom degrees, we deduce the nonlinearity of them and their nonlinear coupling. Through stability analysis, we find that the system presents multi-stability when either the librational or the translational drive is red-detuned. The system will be stabilized if and only if these two drives are both blue-detuned. In the steady state region, we study the synthetic cooling scheme of translational mode by utilising librational mode. We find that matching the driving amplitude of these two modes and appropriate air pressure can optimize synthetic cooling. The synthetic cooling limit can be greatly improved, if we combine the feedback cooling with the synthetic cooling.

\end{abstract}

\maketitle

\section{Introduction}

Advancing research progress in quantum optomechanics has attracted people's a lot interest and paved a way for many applications in the past decade~\cite{KippenbergOe2007,Liu2013,AspelmeyerRmp2014}. The optomechanical systems have been applied in many areas, such as generating macroscopic quantum superpositions and entanglement ~\cite{o2010quantum,chen2013JPB,WangPhysRevA2016}, ultra-sensitive detectors for force~\cite{RanjitPra2015,RanjitPra2016}, quantum information processing \cite{Andrews2014,Yin2015,reed2017faithful}. On the ground of different research demands, people studied different optomechanical systems, such as the microtoroid~\cite{Schliesser2009}, the near-field coupled nano-mechanical oscillators~\cite{AnetsbergerNphys2009}, the membrane \cite{thompson2008strong}, the superconducting circuits~\cite{teufel2011sideband}, the optical levitated nanoparticles~\cite{YinIjmpb2013,neukirch2015nano} and etc. As a novel optomechanical system, the optically levitated system increasingly attracts people's attention due to its high mechanical quality factor Q at vacuum (potentially approaching $Q=10^{12}$) and reconfiguration. Such system can be applied to verify the fundamental principle of quantum mechanics~\cite{Romero-IsartNjp2010,ChangPnas2010,Romero-IsartPrl2011,Yin2017CP} and statistical physics~\cite{LiScience2010,Kheifets2014OScience,Gieseler2014Natnano,Millen2014Natnano,rondin2017Natna,Hoang2018PRL} and to further investigate nonlinear dynamics~\cite{GieselerNphys2013,GieselerPrl2013}, precise measurement and etc~\cite{Geraci2010PRL,LiNphys2011,Yin2011,zhao2014Pra,MoorePrl2014,RashidPrl2016,PhysRevLett.117.163601}.

The optically levitated systems not only have high mechanical Q, but also have multiple mechanical degrees of freedom, such as translation, rotation and libration. The translational mode can be used to measure the instantaneous velocity of Brownian particle~\cite{Kheifets2014OScience,LiScience2010}, the nanoscale temperature \cite{millen2014Natna}, and the magnetic field~\cite{Kumar2017OE}. The rotation of nanorods and nanoparticles levitated by laser beam also attracts a lot of
attentions recently \cite{Kuhn17}. Circularly polarized trapping laser beam is a well adapted method for particle rotation, and the stable rotation rate of particle can reach up to $5~\rm MHz$~\cite{arita2013NatCom} or even $\rm GHz$~\cite{Lukas2018GHz,Ahn2018}. Spatial light modulator based approach~\cite{arita2015OL} and perfect vortex beam with orbital angular momentum~\cite{arita2017JOSAB} also can be utilized for the particle's rotation.  Meanwhile, librational (torsional) mode has been experimentally observed and theoretically explained \cite{VolpePrl2006,PedaciNphys2010,HoangPrl2016}. The sideband cooing scheme of the torsional mode was also proposed ~\cite{HoangPrl2016}. This work stimulated a series of works  such as the decoherence mechanism of the librational modes \cite{Zhong2016,Stickler16a}, and coupling librational modes with the internal spins \cite{Ma2017Pra,Delord2017Pra} or the translational degree of freedom~\cite{liu2017JOSAB}.

While nonlinearity is ubiquitous and could affect the physical property of optomechanics, the nonlinear optomechanical systems can be used to testify phenomenna of fundamental physics~\cite{GieselerNphys2013,GieselerPrl2013,Lv2015}, bistability~\cite{Kyriienko2014PRL,GePra2016,Jiang2016Josab}, multi-stability~\cite{Chang2011Pra} and chaos~\cite{Bakemeier2015PRL,monifi2016optomechanically}. Many applications of nonlinear optomechanical systems have been reported as well, such as ultrasensitivity optical sensor~\cite{FanOe2015,BrawleyNc2016}, cooling by utilising nonlinearity~\cite{FonsecaPrl2016,zhang2017bistable}. In Ref.~\cite{xiao2017Pra}, the nonlinearity of the optically trapped nanoparticle's librational mode is studied. It is found that the red detuning driving can induce bistability of librational mode and single mode squeezing can be realised in blue detuning driving.

Stimulated by the previous investigations on the nonlinearity of the librational mode, here we systematically study the nonlinearity of both the librational and the translational modes of a levitated nonspherical nanoparticle, and derive the nonlinear coupling Hamiltonian between the two modes. The librational and translational modes are stable at the same time if and only if the drives on these two mode are both blue-detuning. As long as the red-detuning drive exists, the two motional modes will be bistable or multi-stable.  In the steady state regime, the linearized beam-splitter Hamiltonian is derived, and the synthetic cooling of the translational mode by the librational mode is studied \cite{PhysRevLett.117.163601}. For this synthetic cooling process, the residual air pressure must be matching with effective coupling strength between the librational and translational modes in the effective beam-splitter like Hamiltonian. Besides, the driving amplitudes of these two modes should match each other in order to obtain the optimal sympathetic cooling ratio. Finally, the feedback cooling can  improve the sympathetic cooling ratio of translational mode.

This paper is organized as follows: In Section II, we theoretically investigate the effective Hamiltonian and the nonlinearity of librational and translational modes of a nonspheric nanoparticle trapped by laser beams. Consequently, in Section III, we take the stability analysis of this nonlinear system. After that in Section IV, we deduce the Beam Splitter Hamiltonian in steady state regime and present that translational mode can be cooled by librational mode. By utilising feedback cooling, we can increase synthetic cooling ratio of the translational mode. Finally in the last section we present a brief conclusion and a perspective to future studies.

\section{The effective Hamiltonian}
\label{Model}
\begin{figure}
\centering
\includegraphics[width=10cm]{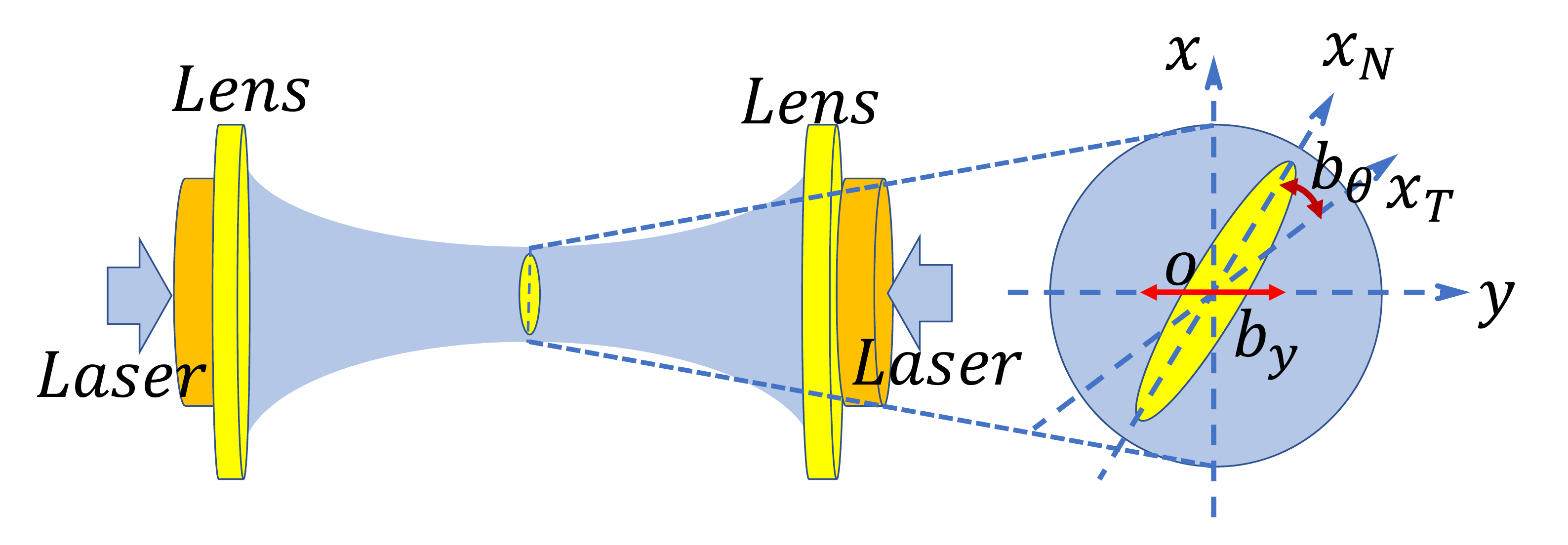}
\caption{(a) A schematic diagram for a ellipsoidal nanoparticle trapped by a laser. The relation between the Cartesian coordinate
systems of the nanoparticle ($x_N$), the trapping laser polarization ($x_T$), and the lens ($x_l$, $y_l$, $z_l$). The $x_N$ axis aligns with
the longest axis of the nanoparticle. $x_T$ and $x_l$ axes align with the trapping laser and the center of the two lenses,
respectively. The angle between $x_N$ and $x_T$ is $\theta$. The librational mode and translational mode are denoted by $b_{\theta}$ and $b_{y}$. The nanoparticle is trapped in a focal plane.}
\label{fig:experiment}
\end{figure}

We consider a nanoparticle that is trapped by two strongly focused laser beams that are counter propagating in horizontal direction as shown in Fig.\ref{fig:experiment}. The nanoparticle has three translational modes and three librational modes in focal plane~\cite{HoangPrl2016,Kuhn17}. Here, we consider one librational mode and one translational mode of an optically levitated ellipsoidal nanoparticle with long axis $r_{a}$, short axis $r_{b}=r_{c}$ and the density $\rho$.
The potential energy of the ellipsoid in the optical tweezers is~\cite{HoangPrl2016}:

\begin{eqnarray}
U(\theta,y)&=&-\frac{V}{2c}[\kappa_{x}-(\kappa_{x}-\kappa_{y})\sin^2\theta]I_{y}(y)\nonumber\\
I_{y}(y)&=&I_{0}e^{-\frac{2y^{2}}{\omega_{0}^{2}}}
\label{Eq:the potential of tweezers}
\end{eqnarray}
where $V=4\pi r_{a}r_{b}^{2}/3$ is the volume of the ellipsoid, $c$ is the speed of the light, $\kappa_{x,y}=\alpha_{x,y}/(\epsilon_{0}V)$ are the effective susceptibility of the ellipsoid, $\epsilon_{0}$ is the vacuum permittivity, and $\theta$ is the angle between the long axis($r_{a}$) of the ellipsoid and the electric field of the trapping laser beam. $I_{y}(y)$ is the intensity of trapping laser along y-direction, $I_{0}$ is the center intensity in the focal plane, $I_{0}=\frac{2P_{0}}{\pi \omega_{0}}^{2}$, $P_{0}$ and $\omega_{0}$ are power and waist of the laser respectively.

The particle tend to minimize its potential energy when it is cooled down. Therefore, both the position $y$ and the angle $\theta$ would approach to zero. Here, in order to obtain the high order
effects of both the librational and translational modes, we expand potential function to the forth order of both $y$ and $\theta$ around the equilibrium position, $y=0$ and $\theta=0$. The potential becomes
\begin{equation}
U(\theta,y)=U_{0}[\kappa_{x}-\frac{2\kappa_x}{\omega_{0}^{2}}y^{2}-\kappa_{xy}\theta^{2}+\frac{2\kappa_x}{\omega_{0}^{4}}y^{4}+\frac{\kappa_{xy}}{3}\theta^{4}+\frac{2\kappa_{xy}}{\omega_{0}^{2}}\theta^{2}y^{2}-\frac{2\kappa_{xy}}{\omega_{0}^{4}}\theta^{2}y^{4}-\frac{2\kappa_{xy}}{3\omega_{0}^{2}}\theta^{4}y^{2}+\frac{2\kappa_{xy}}{3\omega_{0}^{4}}\theta^{4}y^{4}]
\label{the high potential}
\end{equation}
where $U_{0}=-\frac{VI_{0}}{2c}$ and  $\kappa_{xy}=\kappa_{x}-\kappa_{y}$.
In order to quantize the librational mode and translational mode, we define the following operators
 \begin{eqnarray}
\begin{aligned}
\hat{\theta}=\sqrt{\frac{\hbar}{2I\omega_{t}^{\theta}}}(\hat{b}_{\theta}+\hat{b}_{\theta}^{\dag}), \qquad  \hat{p}_{\theta}=i\sqrt{\frac{I\omega_{t}^{\theta}\hbar}{2}}(\hat{b}_{\theta}-\hat{b}_{\theta}^{\dag})\nonumber\\
\hat{y}=\sqrt{\frac{\hbar}{2m\omega_{t}^{y}}}(\hat{b}_{y}+\hat{b}_{y}^{\dag}), \qquad  \hat{p}_{y}=i\sqrt{\frac{m\omega_{t}^{y}\hbar}{2}}(\hat{b}_{y}-\hat{b}_{y}^{\dag}).
\end{aligned}
\label{Eq:b phi y},
\end{eqnarray}
The commutations of them are $[\hat{\theta}, \hat{p}_\theta]=i\hbar$ and $[\hat{y}, \hat{p}_y]= i\hbar$.
The Hamiltonian of this system $H=T+U(\theta,y)$ can be written as
\begin{equation}
 \begin{aligned}
H &=\hbar\omega_{t}^{\theta}\hat{b}^{\dag}_{\theta}\hat{b}_{\theta}+\hbar\omega_{t}^{y}\hat{b}^{\dag}_{y}\hat{b}_{y}-\frac{\hbar^{2}}{8m\omega_{0}^{2}}(\hat{b}_{y}+\hat{b}_{y}^{\dag})^{4}-\frac{\hbar^{2}}{24I}(\hat{b}_{\theta}+\hat{b}_{\theta}^{\dag})^{4}\nonumber\\
&-\frac{\hbar^{2}}{4\omega_{0}I}\sqrt{\frac{(r_{a}^{2}+r_{b}^{2})^{2}\kappa_{xy}}{10\kappa_{x}}}(\hat{b}_{\theta}+\hat{b}_{\theta}^\dag)^{2}(\hat{b}_{y}+\hat{b}_{y}^{\dag})^{2}\nonumber\\
 &+\frac{\hbar^{3}\kappa_{xy}}{16mI\omega_{0}^{2}\kappa_{x}}\sqrt{\frac{c\pi\rho\omega_{0}^{2}(r_{a}^{2}+r_{b}^{2})^{2}}{10\kappa_{xy}P_{0}}}(\hat{b}_{\theta}+\hat{b}_{\theta}^\dag)^{2}(\hat{b}_{y}+\hat{b}_{y}^{\dag})^{4}\\
&+\frac{\hbar^{3}}{48mI}\sqrt{\frac{c\pi\rho}{\kappa_{x}P_{0}}}(\hat{b}_{\theta}+\hat{b}_{\theta}^\dag)^{4}(\hat{b}_{y}+\hat{b}_{y}^{\dag})^{2}
+\frac{\hbar^{4}c\pi\rho}{192m^{2}I\kappa_{x}P_{0}}(\hat{b}_{\theta}+\hat{b}_{\theta}^\dag)^{4}(\hat{b}_{y}+\hat{b}_{y}^{\dag})^{4},
\label{Eq:initial hamiltonian}
\end{aligned}
\end{equation}
where $T=I\dot{\theta}^2/2$, with $I=4\pi\rho r_{a} r_{b}^2(r_{a}^2+r_{b}^2)/15$ being the rotational inertia of the ellipsoid.
In this Hamiltonian, the $4\rm{th}$ order nonlinear coefficient are $\eta_{\theta}=\frac{\hbar}{24I}$, $\eta_{y}=\frac{\hbar}{8m\omega_{0}^{2}}$ and $\eta_{\theta y}=\frac{\hbar}{4\omega_{0}I}\sqrt{\frac{(r_{a}^{2}+r_{b}^{2})^{2}\kappa_{xy}}{10\kappa_{x}}}$, the $6\rm{th}$ order nonlinear coefficients are $\eta_{1}=\frac{\hbar^{2}\kappa_{xy}}{16mI\omega_{0}^{2}\kappa_{x}}\sqrt{\frac{c\pi\rho\omega_{0}^{2}(r_{a}^{2}+r_{b}^{2})^{2}}{10\kappa_{xy}P_{0}}}$ and $\eta_{2}=\frac{\hbar^{2}}{48mI}\sqrt{\frac{c\pi\rho}{\kappa_{x}P_{0}}}$ and the $8\rm{th}$ order nonlinear coefficient is $\eta_{3}=\frac{\hbar^{3}c\pi\rho}{192m^{2}I\kappa_{x}P_{0}}$.

\begin{figure}
\includegraphics[height=2.2in,width=3.5in]{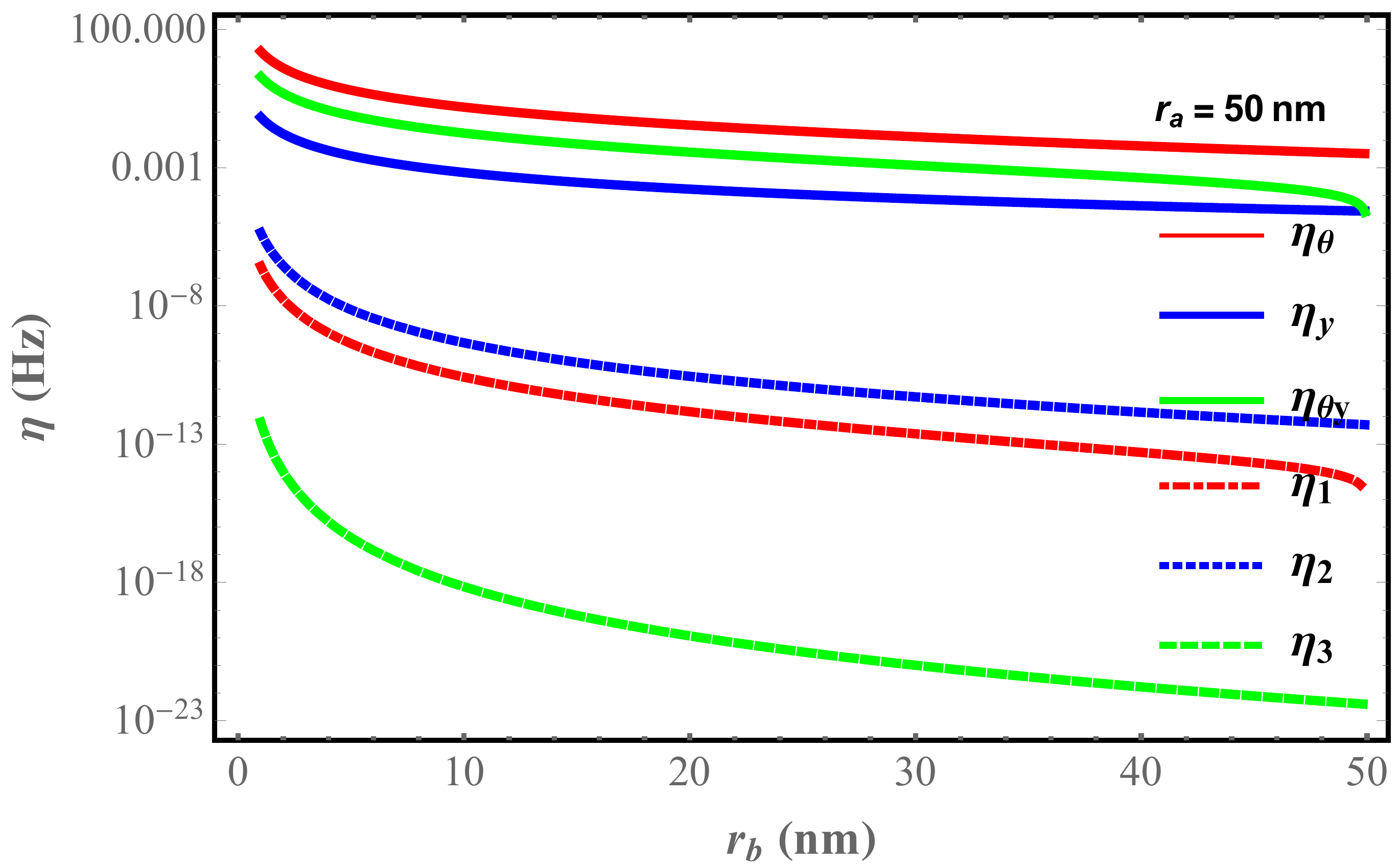}
\caption{The nonlinear coefficients of a glass ellipsoid particle optically levitated by laser beam. Different line presents different order of nonlinearity, where $\eta_{\theta}$, $\eta_{y}$ and $\eta_{\theta y}$ are nonlinearity of librational mode, translational mode and the coupling between them. $\eta_{1}$, $\eta_{2}$ and $\eta_{3}$ are $6\rm th$ and $8\rm th$ order nonlinearity respectively. The $6\rm th$ and $8\rm th$ order nonlinear coefficients can be neglected. The long axis of particle, $r_{a}$, is $50~\rm nm$ and $r_{b}$ is short axis. The laser power and waist are respectively $0.1~\rm W$ and $0.6~\mu m$.}
\label{fig:nonlinear coefficients}
\end{figure}

As shown in Fig.~\ref{fig:nonlinear coefficients}, both
 the $8\rm{th}$ order and the $6\rm{th}$ order terms are much less than the $4\rm{th}$ order term. Therefore, the $6\rm{th}$ and $8\rm{th}$ order nonlinear terms can be omitted.
 The Hamiltonian Eq.(~\ref{Eq:initial hamiltonian}) can be simplified as follow,
\begin{eqnarray}
\hat{H}_{r}&=&\hbar\omega_{t}^{\theta}\hat{b}^{\dag}_{\theta}\hat{b}_{\theta}+\hbar\omega_{t}^{y}\hat{b}^{\dag}_{y}\hat{b}_{y}-\frac{\hbar^{2}}{8m\omega_{0}^{2}}(\hat{b}_{y}+\hat{b}_{y}^{\dag})^{4}-\frac{\hbar^{2}}{24I}(\hat{b}_{\theta}+\hat{b}_{\theta}^{\dag})^{4}\nonumber\\
& &-\frac{\hbar^{2}}{4\omega_{0}I}\sqrt{\frac{(r_{a}^{2}+r_{b}^{2})^{2}\kappa_{xy}}{10\kappa_{x}}}(\hat{b}_{\theta}+\hat{b}_{\theta}^\dag)^{2}(\hat{b}_{y}+\hat{b}_{y}^{\dag})^{2}
\label{Eq:reduced Hamiltonian}.
\end{eqnarray}
Both librational and translational modes can be driven by lasers, whose driving amplitudes and frequencies are respectively $\Omega_{1}$, $\Omega_{2}$, $\omega_{l1}$ and $\omega_{l2}$ for both librational and translational modes~\cite{xiao2017Pra}, the driving Hamiltonian can be described as
\begin{equation}
\hat{H}_{dr}=\frac{\hbar\Omega_{1}}{2}(\hat{b}_{\theta}e^{\rm{i}\omega_{l1}t}+\hat{b}_{\theta}^{\dag}e^{-\rm{i}\omega_{l1}t})+\frac{\hbar\Omega_{2}}{2}(\hat{b}_{y}e^{\rm{i}\omega_{l2}t}+\hat{b}_{y}^{\dag}e^{-\rm{i}\omega_{l2}t})
\label{Eq:driven Hamiltonian}.
\end{equation}

In rotating wave frame, the Hamiltonian is transformed following $\hat{H}_{RM}=\hat{U}^{\dag}(\hat{H}_{r}+\hat{H}_{dr})\hat{U}-\hbar\omega_{l1}\hat{b}_{\theta}^{\dag}\hat{b}_{\theta}-\hbar\omega_{l2}\hat{b}_{y}^{\dag}\hat{b}_{y}$ and $\hat{U}=\mathrm{e}^{-\mathrm{i}(\omega_{l1}\hat{b}_{\theta}^{\dag}\hat{b}_{\theta}t+\omega_{l2}\hat{b}_{y}^{\dag}\hat{b}_{y}t)}$, and rotating wave approximation can be utilized for this system. Under the condition $\omega_{l1}\neq\omega_{l2}$, the effective Hamiltonian can be written as
\begin{eqnarray}
\hat{H}_{RWA}=&-&\hbar(\Delta_{l1}+2\eta_{\theta y})\hat{b}_{\theta}^{\dag}\hat{b}_{\theta}-\hbar(\Delta_{l2}+2\eta_{\theta y})\hat{b}_{y}^{\dag}\hat{b}_{y}+\frac{\hbar\Omega_{1}}{2}(\hat{b}_{\theta}+\hat{b}_{\theta}^{\dag})+\frac{\hbar\Omega_{2}}{2}(\hat{b}_{y}+\hat{b}_{y}^{\dag})\nonumber\\
&-&3\hbar\eta_{\theta}(\hat{b}_{\theta}^{\dag 2}\hat{b}_{\theta}^{2}+\hat{b}_{\theta}^{2}\hat{b}_{\theta}^{\dag 2})-3\hbar\eta_{y}(\hat{b}_{y}^{\dag 2}\hat{b}_{y}^{2}+\hat{b}_{y}^{2}\hat{b}_{y}^{\dag 2})-4\hbar\eta_{\theta y}\hat{b}_{\theta}^{\dag}\hat{b}_{\theta}\hat{b}_{y}^{\dag}\hat{b}_{y}
\label{Eq:Hamiltonian in RW Approximation}.
\end{eqnarray}
Here we neglect the highly oscillation terms with frequency of $\pm2\omega_{l1/l2}$ and $\pm4\omega_{l1/l2}$. $\Delta_{l1/l2}=\omega_{l1/l2}-\omega_{t}^{\theta/y}$.
The above Hamiltonian contains the nonlinear terms for both the librational and the translational modes, and the nonlinear coupling between them.

\section{Multi-stability, Bistability and stable conditions}
\label{coupling_bistability}
In previous section, we have deduced an effective Hamiltonian \eqref{Eq:Hamiltonian in RW Approximation} for the system with high nonlinearity.
The system may show bistable, even multi-stable states other than stable states through the specific drivings.
We will study the stable condition through master equation method based on Hamiltonian \eqref{Eq:Hamiltonian in RW Approximation} in this section.
The master equation that describes the dynamics of a nanoparticle which couples with the thermal bath~\cite{Louisell1973quantum} is
\begin{eqnarray}
\dot{\hat{\rho}}(t)=\frac{1}{\mathrm{i}\hbar}[\hat{H}_{RWA}(t),\hat{\rho}]+\mathscr{L}_{\theta}\hat{\rho}+\mathscr{L}_{y}\hat{\rho},
\label{Eq:master equation},
\end{eqnarray}
where $\mathscr{L}_{\theta/y}=\frac{(1+\overline{n}_{\theta/y})}{2}\gamma_{\theta/y}\mathscr{D}_{b_{\theta/y}}+\frac{\overline{n}_{\theta/y}}{2}\gamma_{\theta/y}\mathscr{D}_{b_{\theta/y}^{\dag}}$, with the Lindblad operator $\mathscr{D}_{x}(\rho)=2x\rho x^{\dag}-x^{\dag}x\rho-\rho x^{\dag}x$. Here $\gamma_{\theta/y}$ is the decay rate of librational (translational) mode, and $\overline{n}_{\theta/y}$ is the average phonon number of the librational (translational) thermal reservoir. To investigate the steady state property and the quantum fluctuation of both the librational and the translational modes, the amplitude of the librational mode $\beta_{\theta/y}(t)$ is split into two terms: the average amplitude $\beta_{\theta/y}$ and fluctuation $\delta b_{\theta/y}(t)$. Using the master Eq.(~\ref{Eq:master equation}), we can deduce that the motional equations for $\beta_{\theta}$ and $\beta_{y}$
\begin{equation}
\frac{\partial}{\partial t}
\begin{pmatrix}
\beta_{\theta}\\ \beta_{y}
\end{pmatrix}=
\begin{pmatrix}
\Big(-\frac{\gamma_{\theta}}{2}-\mathrm{i}\big(\Delta_{l1}-12\eta_{\theta}(|\beta_{\theta}|^{2}+1)-4\eta_{\theta y}|\beta_{y}|^{2}\big)\Big)\beta_{\theta}-\mathrm{i}\frac{\Omega_{1}}{2}\\
\Big(-\frac{\gamma_{y}}{2}-\mathrm{i}\big(\Delta_{l2}-12\eta_{y}(|\beta_{y}|^{2}+1)-4\eta_{\theta y}|\beta_{\theta}|^{2}\big)\Big)\beta_{y}-\mathrm{i}\frac{\Omega_{2}}{2}
\end{pmatrix}.
\label{Eq:matrix of beta}
\end{equation}
Here we apply the semiclassical approximation, and neglect the terms $\left\langle\hat{b}_{\theta/y}^{\dag}\hat{b}_{\theta/y}^{2}\right\rangle-\left\langle \hat{b}_{\theta/y}^{\dag}\right\rangle\left\langle\hat{b}_{\theta/y}^{2}\right\rangle$. Therefore, we have $\left\langle \hat{b}_{\theta/y}^{\dag}\hat{b}_{\theta/y}^{2}\right\rangle=\beta_{\theta/y}^{\ast}\beta_{\theta/y}^{2}$. This approximation requires fluctuation terms $\left\langle(\delta\hat{b}_{\theta/y})^{2}\right\rangle$ and $\left\langle\delta\hat{b}_{\theta/y}^{\dag}\delta\hat{b}_{\theta/y}\right\rangle$ to be much less than $|\beta_{\theta/y}|^2$.

\begin{figure}[!b]
\centering
\subfigure{
\begin{minipage}{0.43\textwidth}
\includegraphics[width=1\textwidth]{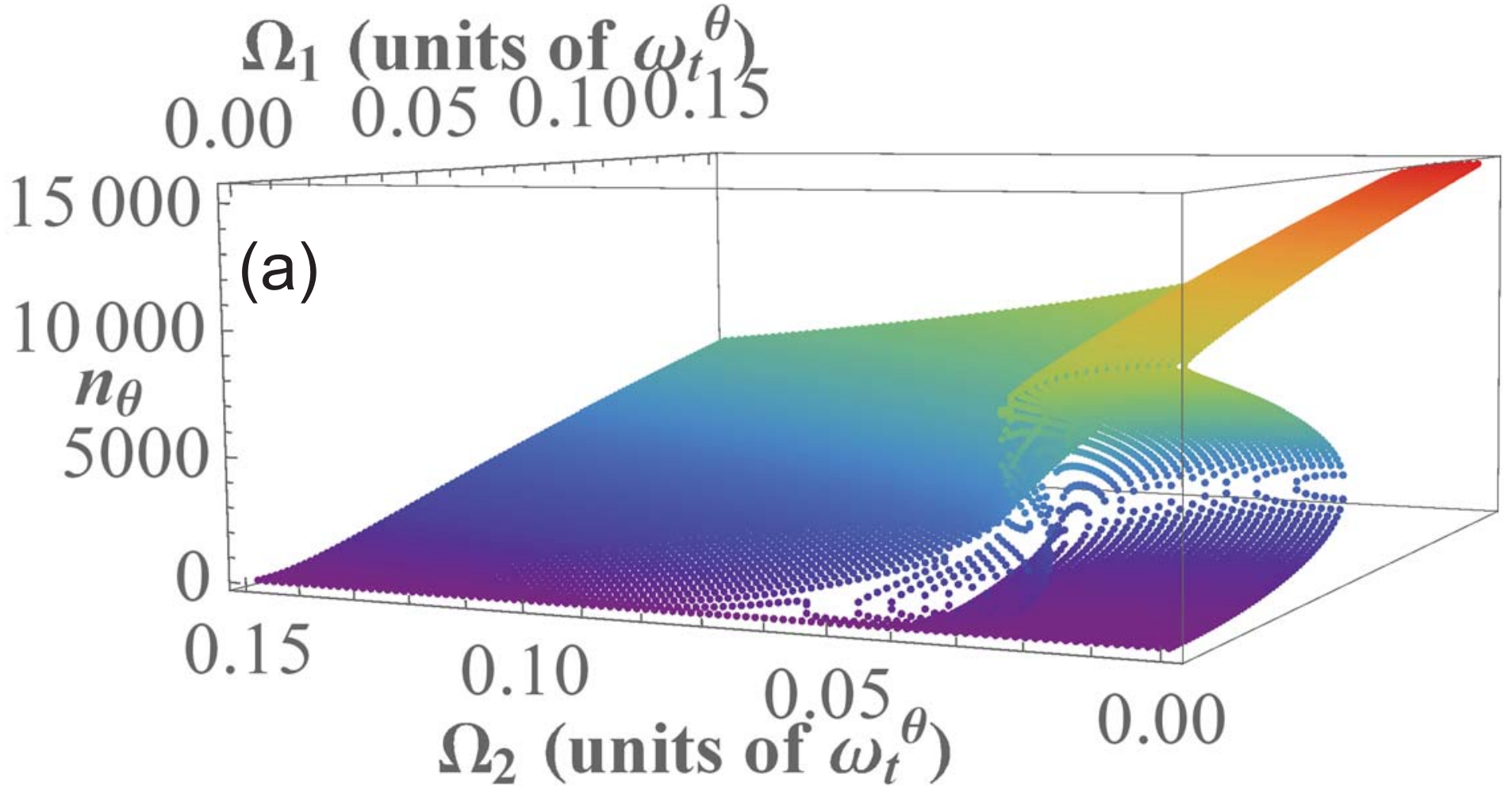}
\label{fig:multistabilityless1lib3d}
\end{minipage}}
\subfigure{
\begin{minipage}{0.48\textwidth}
\includegraphics[width=1\textwidth]{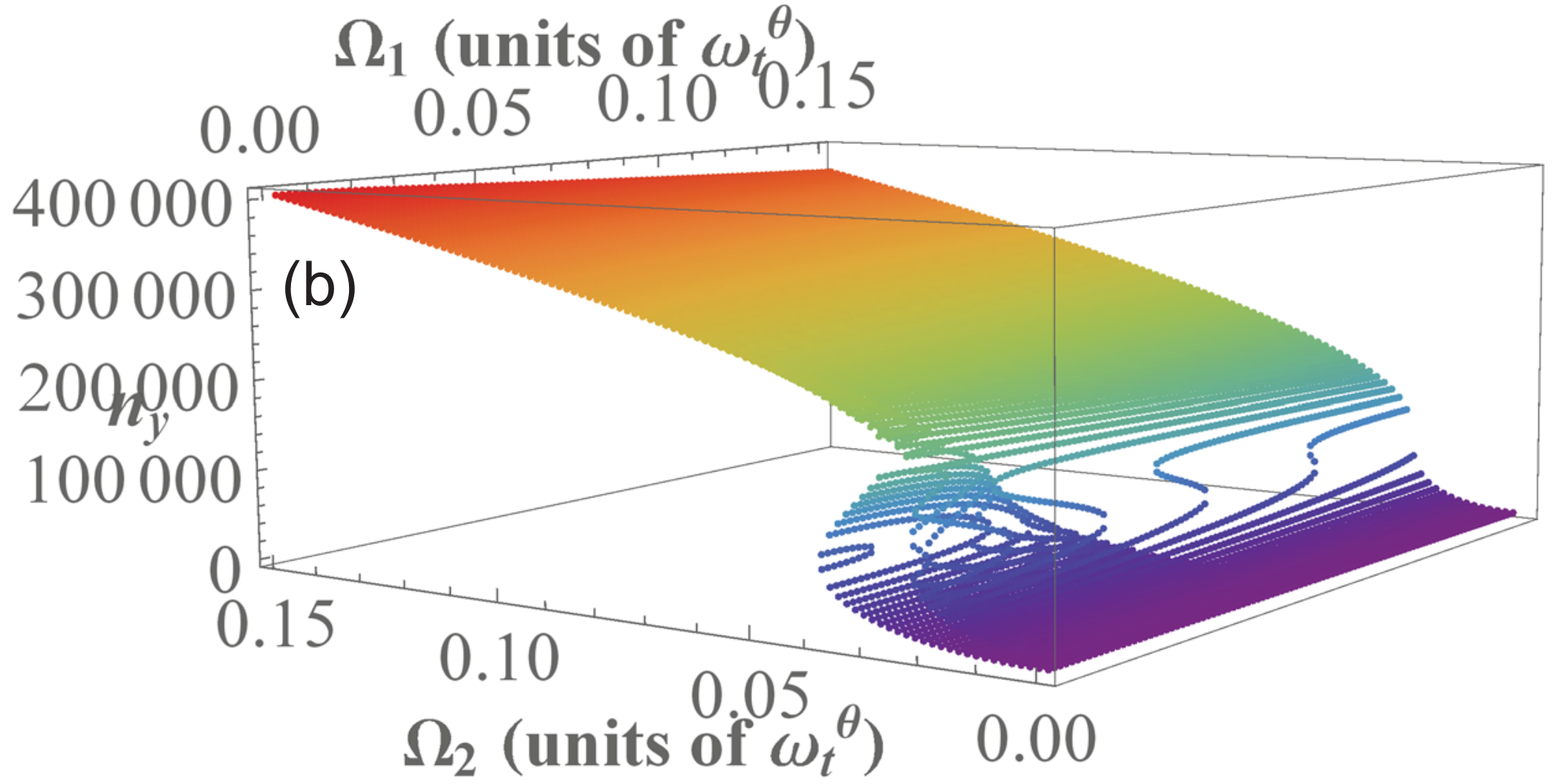}
\label{fig:multistabilityless1com3d}
\end{minipage}}
 \caption{Multi-stability of librational mode and translational mode in this system when $\Delta_{1}=0.01\omega_{t}^{\theta}$ and $\Delta_{2}=0.01\omega_{t}^{y}$. (a) the average phonon number of librational mode $\tilde{n}_{\theta}$ as a function of $\Omega_{1}$ and $\Omega_{2}$ in the red detuning drive. (b) the average phonon number of translational mode $\tilde{n}_{y}$ as a function of $\Omega_{1}$ and $\Omega_{2}$ in the red detuning drive where the residual air pressure $P=1~\rm{mPa}$ and temperature $T=300~\rm{K}$. The long axis and short axis are respectively $50~\rm{nm}$ and $25~\rm{nm}$, $\Omega_{1}$ and $\Omega_{2}$ are in units of $\omega_{t}^{\theta}$.}
\label{fig:multistability in red detuning of two drive}
\end{figure}

To study the steady state more precisely, we derive the equations for the steady state as follows:
 \begin{eqnarray}
   \Big(-\frac{\gamma_{\theta}^{2}}{4}+\big(\Delta_{1}-12\eta_{\theta}(\tilde{n}_{\theta}+1)-4\eta_{\theta y}\tilde{n}_{y}\big)^{2}\Big)\tilde{n}_{\theta}&=&\frac{\Omega_{1}^{2}}{4}, \nonumber\\
   \Big(-\frac{\gamma_{y}^{2}}{4}+\big(\Delta_{2}-12\eta_{y}(\tilde{n}_{y}+1)-4\eta_{\theta y}\tilde{n}_{\theta}\big)^{2}\Big)\tilde{n}_{y}&=&\frac{\Omega_{2}^{2}}{4},
   \label{Eq:steady state equation}
 \end{eqnarray}
where $\tilde{n}_{\theta/y}=|\beta_{\theta/y}|^{2}$ is the average phonon number of the librational (translational) mode and $\Delta_{1/2}=\omega_{t}^{\theta/y}-\omega_{l1/l2}-2\eta_{\theta y}$. Here we set all parameters ($\gamma_{\theta}, \gamma_{y}, \Delta_{1}, \Delta_{2}, \eta_{\theta}, \eta_{y}, \eta_{\theta y}, \Omega_{1}, \Omega_{2}$) are in units of $\omega_{t}^{\theta}$.
The analytical approach to such equations is still under developed, so we study the system through numerical method. In Eq.~\ref{Eq:steady state equation}, both detuning $\Delta_{1}$ and $\Delta_{2}$ influence the steady state.  Different detuning could give various steady state property. For example, when $\Delta_{1}$ and $\Delta_{2}$ are larger than zero, in another word, where both drives are red detuning, the system presents multi-stability property. As shown in Fig.~\ref{fig:multistability in red detuning of two drive}, both $\tilde{n}_{\theta}$ and $\tilde{n}_{y}$ show multi-stability in some parameters region. In this example, we choose an ellipsoidal glass particle with long axis $r_{a}=50~\rm{nm}$ and short axis $r_{b}=25~\rm{nm}$ is trapped by laser, whose power $P=0.1~\rm{W}$ and waist $w=0.6~\mu \rm{m}$, therefore, the oscillating frequency of librational mode and translational mode are respectively $\omega_{t}^{\theta}=2.34~\rm{MHz}$ and $\omega_{t}^{y}=24.5~\rm{kHz}$. The pressure of the residual air  $P=10^{-3}~\rm{Pa}$ and the temperature $T=300~\rm{K}$. Hence the damping of librational mode and translational mode, $\gamma_{\theta}=137.2~\rm{Hz}$ and $\gamma_{y}=47~\rm{Hz}$. The nonlinear coefficients, $\eta_{\theta}=0.202~\rm{Hz}$, $\eta_{y}=0.105~\rm{mHz}$ and $\eta_{\theta y}=2.01~\rm{mHz}$.
Here we suppose that the driving frequencies are fixed while the driving amplitudes are changeable. The driving frequencies are red detuning and we have taken more examples for investigations of stability in Appendix \ref{SectVI}. We find that the red detuning drive will make system present bistability even multi-stability. Therefore, we should find the relation between average phonon number of librational mode (translational mode) and driving frequencies in order to find the steady state region of this system under the condition of fixed driving amplitudes.

From the above example, we find that
the driving frequencies determine the stability of the librational and translational modes~\cite{xiao2017Pra}. If the driving amplitudes of both the librational and translational modes are fixed,  we can tune the driving frequencies to study the stability of these modes. In this case, we set the drive amplitudes $\Omega_{1}=0.5$ and $\Omega_{2}=0.5$.
According to the numerical simulation of Eq.(\ref{Eq:steady state equation}), the detunings $\Delta_{1}$ and $\Delta_{2}$ can affect the average phonon number of librational ($n_{\theta}$) and translational ($n_{y}$) modes. 
\begin{figure}[!b]
\centering
\subfigure{
\begin{minipage}{0.43\textwidth}
\includegraphics[width=1\textwidth]{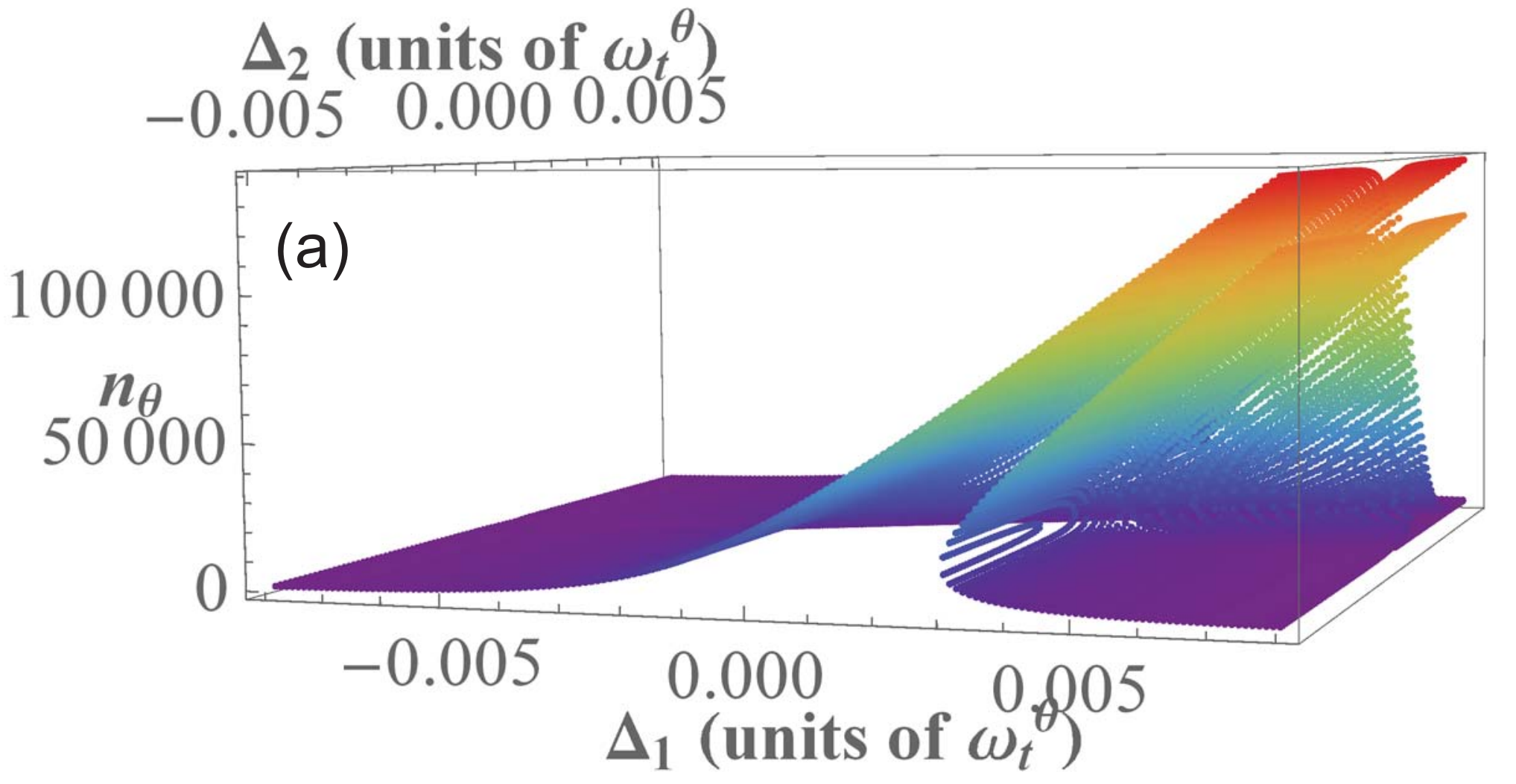}
\label{fig:multistabilitybluedetuninglib}
\end{minipage}}
\subfigure{
\begin{minipage}{0.45\textwidth}
\includegraphics[width=1\textwidth]{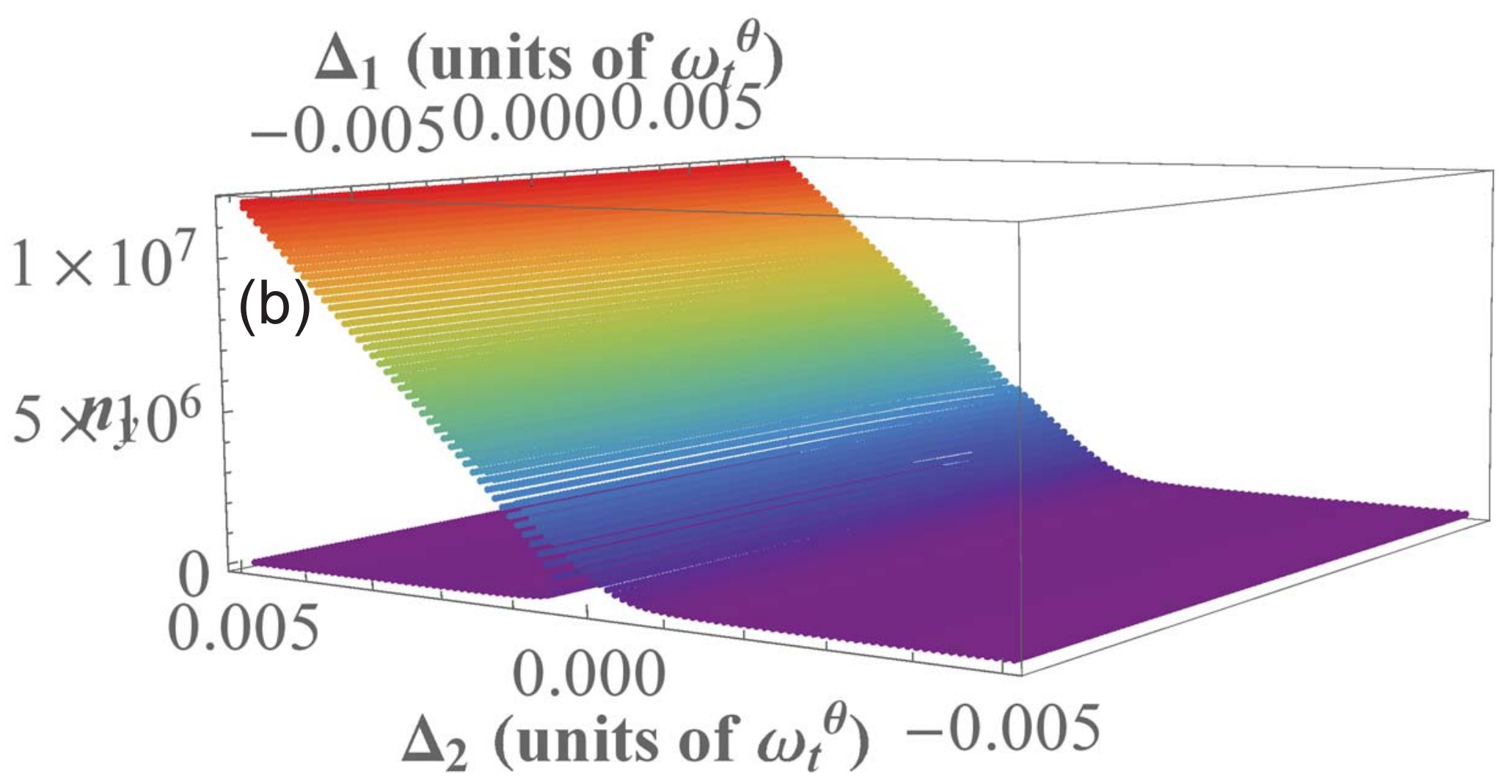}
\label{fig:multistabilitybluedetuningcom}
\end{minipage}}
 \caption{Multi-stability and Bistability of librational mode and translational mode driven by amplitudes $\Omega_{1}=0.5$ and $\Omega_{2}=0.5$. The average phonon number of librational mode $\tilde{n}_{\theta}$ (a) and the average phonon number of translational mode $\tilde{n}_{y}$ (b) are dependent of $\Delta_{1}$ and $\Delta_{2}$. Other parameters are the same to those in Fig.\ref{fig:multistability in red detuning of two drive}.}
\label{fig:frequency bistability}
\end{figure}
As shown in Fig.~\ref{fig:frequency bistability}, if the effective detuning of the librational mode and translational mode are both less than zero, ($\Delta_{1}<0$ and $\Delta_{2}<0$ ), they will both be in steady state. When $\Delta_{2}<0$ and $\Delta_{1}$ is arbitrary, the average phonon number of translational mode will be in steady state. For the rest cases the librational mode and translational mode will not present steady state. In conclusion, the driving frequency determines the stability of librational and translational modes. If both the librational and translational modes are simultaneously in the steady state, the conditions $\Delta_{1}<0$ and $\Delta_{2}<0$ should be satisfied.

\begin{figure}[!b]
\centering
\subfigure{
\begin{minipage}{0.45\textwidth}
\includegraphics[width=1\textwidth]{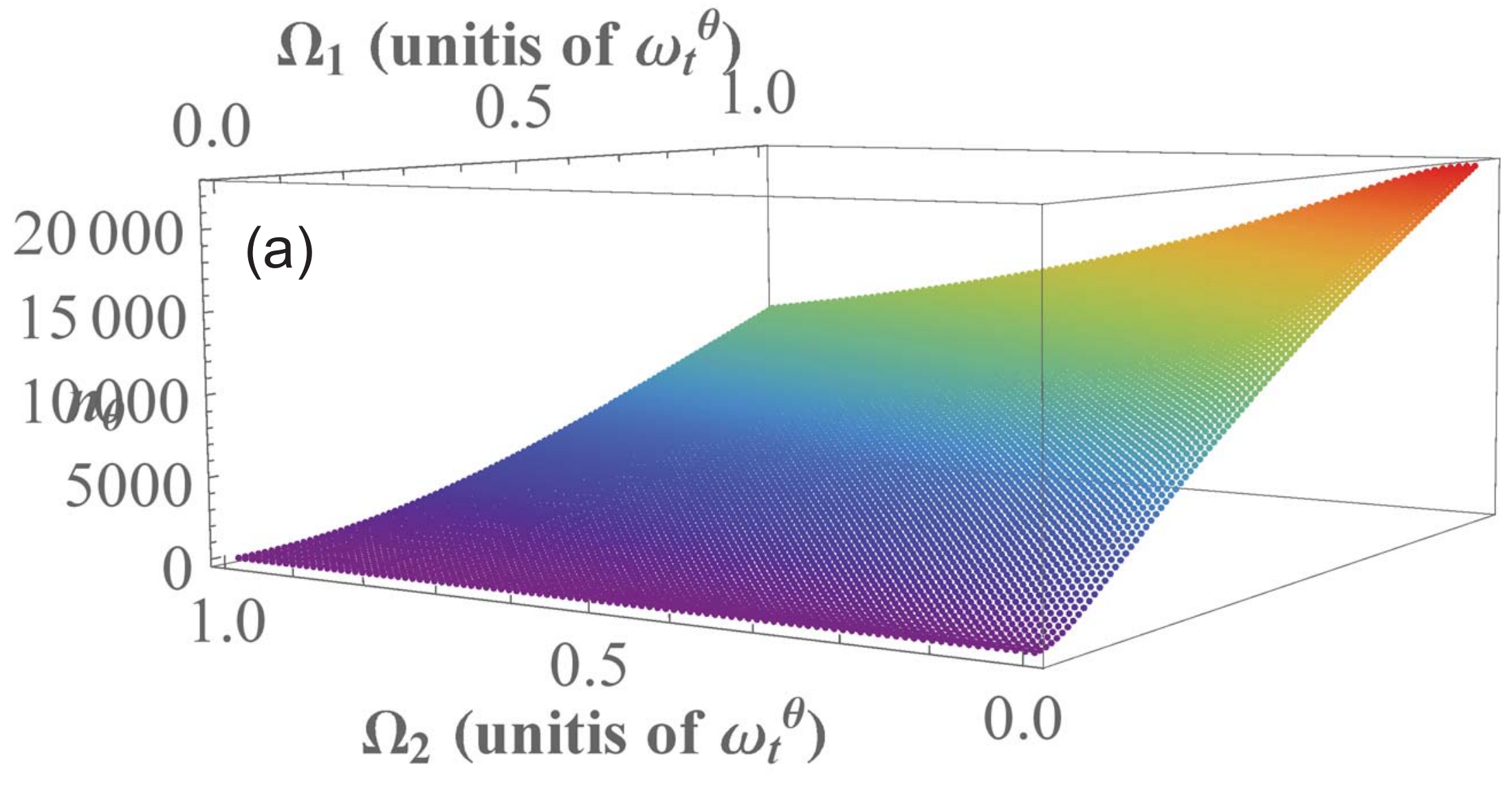}
\label{fig:multistabilitybluedetuninglib1}
\end{minipage}}
\subfigure{
\begin{minipage}{0.45\textwidth}
\includegraphics[width=1\textwidth]{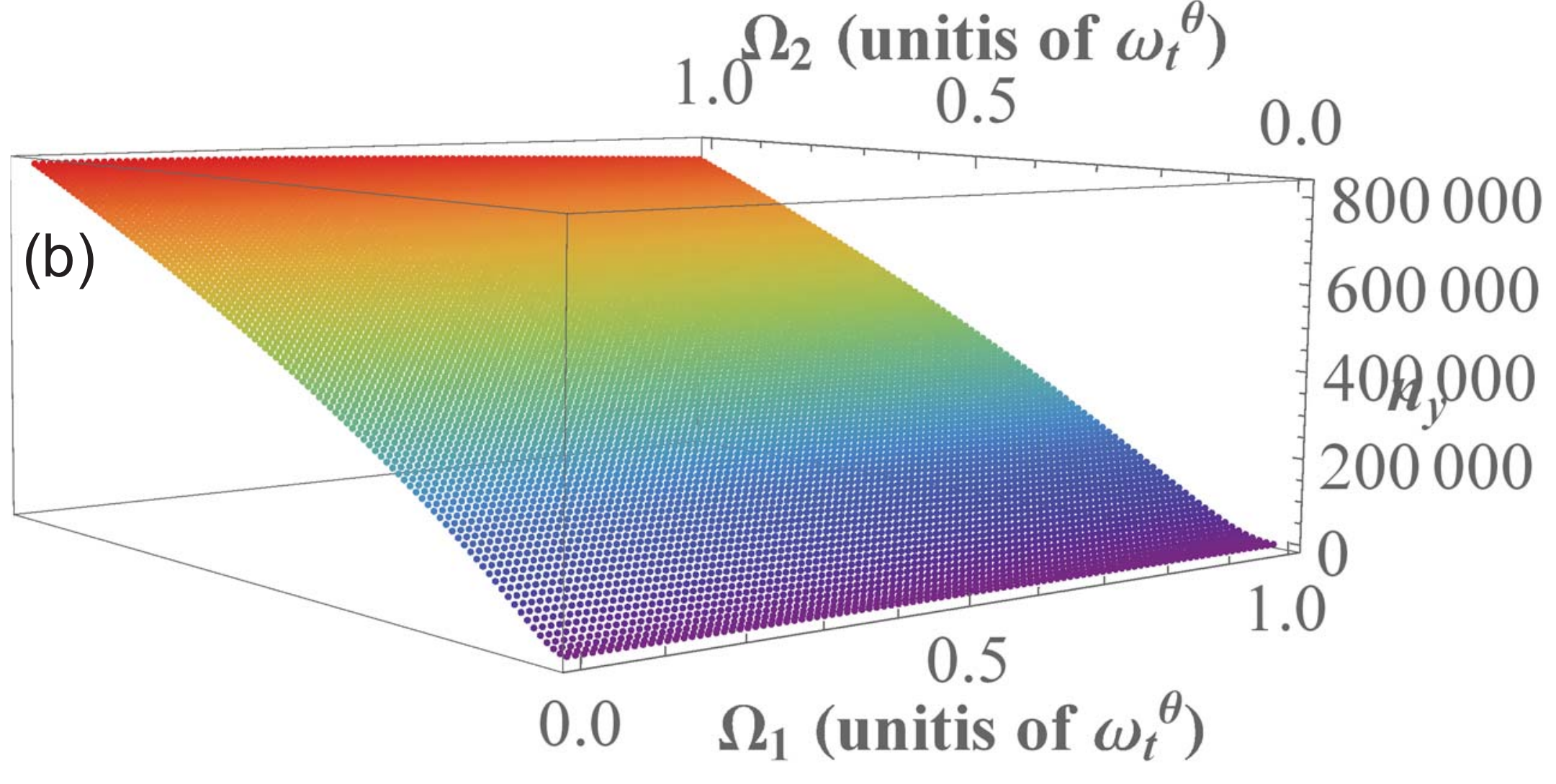}
\label{fig:multistabilitybluedetuningcom1}
\end{minipage}}
 \caption{Steady state of librational mode and translational mode in $\Delta_{1}=-0.1$ and $\Delta_{2}=-0.1\omega_{t}^{y}$ this system. (a) the average phonon number of librational mode $\tilde{n}_{\theta}$ versus $\Omega_{1}$ and $\Omega_{2}$. (b) the average phonon number of translational mode $\tilde{n}_{y}$ versus $\Omega_{1}$ and $\Omega_{2}$. Other parameters are the same to Fig.\ref{fig:multistability in red detuning of two drive}.}
\label{fig:single steady state in bule detuning}
\end{figure}

Another steady state example for the $\tilde{n}_{\theta}$ and $\tilde{n}_{y}$ is shown in Fig.~\ref{fig:single steady state in bule detuning}, where both two modes are in the blue detuning drives.
In this case, the average phonon number $\tilde{n}_{\theta}$ and $\tilde{n}_{y}$ can present steady state property in arbitrary drive amplitudes $\Omega_{1}$ and $\Omega_{2}$. This example verifies the conclusion that the coupling bistable system have steady state in blue detuning drivings for arbitrary driving amplitudes.

 In summary, $\tilde{n}_{\theta}$ and $\tilde{n}_{y}$ have steady state, bistable state and multistable state in different parameters region shown as above. When one of the drive frequencies $\omega_{l1}$ and $\omega_{l2}$ is red detuning, the average phonon number $\tilde{n}_{\theta}$ and $\tilde{n}_{y}$ will show multi-stability or bistablility in small drive amplitude, however the increment of drive amplitude can make $\tilde{n}_{\theta}$ and $\tilde{y}$ present steady state. It is interesting that when $\omega_{l1}$ and $\omega_{l2}$ are blue detuning, the average phonon number $\tilde{n}_{\theta}$ and $\tilde{n}_{y}$ always present steady state for arbitrary drive amplitude. When the system is in the steady state regime, the driving induced effective coupling between the librational and translational modes can be useful for synthetic cooling.

\section{Synthetic cooling of  translational mode}
\label{SectIV}
In the previous section, we discussed the multi-stability of the librational and translational modes, and found the parameter region for the steady states.
As we know, the nonlinearity not only stimulates the multi-stability but also induces some novel quantum properties. Here, we discuss another nonlinearity induced phenomena, the synthetic cooling of the translational mode by the librational mode. Through the standard linearization method, we can get the linearized effective Hamiltonian and set $\delta \hat{b}_{\theta/y}\to\hat{b}_{\theta/y}$ for simplicity, and $|\beta_{\theta/y}|\gg 1$, the linearized Hamiltonian reads
\begin{eqnarray}
\hat{H}_{l}&=&\hat{H}_{0l}+\hat{H}_{\theta l}+\hat{H}_{y l}+\hat{H}_{\theta l}
\label{Eq:linearisated Hamiltonian1}
\end{eqnarray}
where
\begin{eqnarray}
\hat{H}_{0l}&=&\hbar\big(\Delta_{1}-24\eta_{\theta}|\beta_{\theta}|^{2}-4\eta_{\theta y}(|\beta_{y}|^{2}+1)\big)\hat{b}_{\theta}^{\dag}\hat{b}_{\theta}+\hbar\big(\Delta_{2}-24\eta_{y}|\beta_{y}|^{2}-4\eta_{\theta y}(|\beta_{\theta}|^{2}+1)\big)\hat{b}_{y}^{\dag}\hat{b}_{y},\nonumber\\
\hat{H}_{\theta l}&=&-3\hbar\eta_{\theta}(\hat{b}_{\theta}^{\dag 2}\hat{b}_{\theta}^{2}+2\beta_{\theta}^{2}\hat{b}_{\theta}^{\dag 2}+2\beta_{\theta}\hat{b}_{\theta}^{\dag 2}\hat{b}_{\theta}+2\beta_{\theta}\hat{b}_{\theta}\hat{b}_{\theta}^{\dag 2}+\rm{H.C.}),\\
\hat{H}_{y l}&=&-3\hbar\eta_{y}(\hat{b}_{y}^{\dag 2}\hat{b}_{y}^{2}+2\beta_{y}^{2}\hat{b}_{y}^{\dag 2}+2\beta_{y}\hat{b}_{y}^{\dag 2}\hat{b}_{y}+2\beta_{y}\hat{b}_{y}\hat{b}_{y}^{\dag 2}+\rm{H.C.}),\nonumber\\
\hat{H}_{\theta y}&=&-4\hbar\eta_{\theta y}(\frac{1}{2}\hat{b}_{\theta}^{\dag}\hat{b}_{\theta}\hat{b}_{y}^{\dag}\hat{b}_{y}+\beta_{y}\hat{b}_{\theta}\hat{b}_{y}^{\dag}\hat{b}_{\theta}+\beta_{\theta}\hat{b}_{\theta}^{\dag}\hat{b}_{y}^{\dag}\hat{b}_{y}+\beta_{\theta}\beta_{y}\hat{b}_{\theta}^{\dag}\hat{b}_{y}^{\dag}+\beta_{\theta}\beta_{y}^{\ast}\hat{b}_{\theta}^{\dag}\hat{b}_{y}+\rm{H.C.}).\nonumber
\label{Eq:linearisated Hamiltonian for every system}
\end{eqnarray}

It is clear that the condition $\Delta_{1}-24\eta_{\theta}|\beta_{\theta}|^{2}-4\eta_{\theta y}(|\beta_{y}|^{2}+1)=\Delta_{2}-24\eta_{y}|\beta_{y}|^{2}-4\eta_{\theta y}(|\beta_{\theta}|^{2}+1)+\delta$ can be satisfied by adjusting the drive frequencies $\omega_{l1}$ and $\omega_{l2}$. For convenience, $\Delta_{{\rm eff}1}=\Delta_{1}-24\eta_{\theta}|\beta_{\theta}|^{2}-4\eta_{\theta y}(|\beta_{y}|^{2}+1)$ and $\Delta_{{\rm eff}}=\Delta_{2}-24\eta_{y}|\beta_{y}|^{2}-4\eta_{\theta y}(|\beta_{\theta}|^{2}+1)$ are fixed, and $\Delta_{{\rm eff}1}-\Delta_{{\rm eff}2}=\delta$. By utilizing the rotating frame Transformation and the rotating wave approximation, Eq.(\ref{Eq:linearisated Hamiltonian1}) can be transformed into beam-splitter-like Hamiltonian
\begin{equation}
\hat{H}_{\rm{bs}}=\hat{H}_{\rm{bs}1}+\hat{H}_{\rm{bs}2}=-\hbar\delta\hat{b}_{y}^{\dag}\hat{b}_{y}-4\hbar\eta_{\theta y}(\beta_{\theta}\beta_{y}^{\ast}\hat{b}_{\theta}^{\dag}\hat{b}_{y}+{\rm H.C.}),
\label{Eq:beam spliter Hamiltonian}
\end{equation}
where $\hat{H}_{\rm{bs}1}=-\hbar\delta\hat{b}_{y}^{\dag}\hat{b}_{y}$, and $\hat{H}_{\rm{bs}2}=-4\hbar\eta_{\theta y}(\beta_{\theta}\beta_{y}^{\ast}\hat{b}_{\theta}^{\dag}\hat{b}_{y}+{\rm H.C.})$.

Based on Hamiltonian \eqref{Eq:beam spliter Hamiltonian}, the master equation of the system is
\begin{equation}
\dot{\hat{\rho}}(t)=\frac{1}{\mathrm{i}\hbar}[\hat{H}_{\rm{bs}}(t),\hat{\rho}]+\mathscr{L}_{\theta}\hat{\rho}+\mathscr{L}_{y}\hat{\rho},
\label{Eq:beam spliter master equation}
\end{equation}
where $\mathscr{L}_{\theta} =\frac{\gamma_{\theta}}{2}(1+\bar{n}_{\theta})\mathscr{D}_{\theta}+\frac{\gamma_{\theta}}{2}(\bar{n}_{\theta})\mathscr{D}_{\theta^{\dag}}$, $
\mathscr{L}_{y}=\frac{\gamma_{y}}{2}(1+\bar{n}_{y})\mathscr{D}_{y}+\frac{\gamma_{y}}{2}(\bar{n}_{y})\mathscr{D}_{y^{\dag}}$
and $\mathscr{D}_{x}=2x\rho x^{\dag}-x^{\dag}x\rho-\rho x^{\dag}x$ is the Lindblad superoperation for $x$ to be $\theta$ or $y$. $\overline{n}_{\theta}$ and $\overline{n}_{y}$ are the average thermal phonon number of librational and translational mode reservoir, respectively. From Eq.(~\ref{Eq:beam spliter master equation}), we can adiabatically eliminated the librational mode to get the reduced master equation for the translational mode, and vise versa. Therefore, we can define two superoperators $\mathscr{L}_{int}$ and $\mathscr{L}_{free}$.
\begin{eqnarray}
\mathscr{L}_{int}&=&-\frac{i}{\hbar}[\hat{H}_{\rm{bs1}},\cdot], \nonumber\\
\mathscr{L}_{free}&=&-\frac{i}{\hbar}[\hat{H}_{\rm{bs2}},\cdot]+\mathscr{L}_{\theta}+\mathscr{L}_{y}.
\label{Eq:new operator for reduce masterequa}
\end{eqnarray}
In the weak coupling limit, we get the reduced density matrix $\rho_{y}$ which satisfies
\begin{equation}
\dot{\rho}_{y}(t)=-\frac{i}{\hbar}[\hat{H}_{{\rm bs}1},\rho_{y}(t)]+\mathscr{L}_{y}-\frac{1}{\hbar^{2}}\rm{Tr}_{\theta}\Big(\Big[\hat{H}_{\rm{bs}2},\int_{0}^{\infty}\rm{d}t^{'}[e^{\mathscr{L}_{free}t^{'}}(\hat{H}_{\rm{bs}2}),\rho_{y}(t)\bigotimes\rho_{\theta}]\Big]\Big).
\label{Eq:reduced master equation about b_y simplified}
\end{equation}
Therefore, we can have
\begin{eqnarray}
\frac{\rm{d}}{\rm{d}t}\rho_{y}(t)&=&i\delta\Big(1-\frac{64\eta_{\theta y}^{2}|\beta_{\theta}\beta_{y}|^{2}}{(\gamma_{\theta}+\gamma_{y})^{2}+4\delta^{2}}\Big)[\hat{b}_{y}^{\dag}\hat{b}_{y},\rho_{y}]\nonumber\\
& &+\Big(\frac{\gamma_{y}}{2}(1+\bar{n}_{y})+\frac{32(\gamma_{\theta}+\gamma_{y})\eta_{\theta y}^{2}|\beta_{\theta}\beta_{y}|^{2}}{(\gamma_{\theta}+\gamma_{y})^{2}+4\delta^{2}}(1+\langle\hat{n}_{\theta}\rangle)\Big)\mathscr{D}_{y}(\rho_{y}(t))\\
& &\Big(\frac{\gamma_{y}}{2}\bar{n}_{y}+\frac{32(\gamma_{\theta}+\gamma_{y})\eta_{\theta y}^{2}|\beta_{\theta}\beta_{y}|^{2}}{(\gamma_{\theta}+\gamma_{y})^{2}+4\delta^{2}}\langle\hat{n}_{\theta}\rangle\Big)\mathscr{D}_{y^{\dag}}(\rho_{y}(t))\nonumber
\label{Eq:dot rhot}.
\end{eqnarray}
By setting $\tilde{\eta}=\frac{32(\gamma_{\theta}+\gamma_{y})\eta_{\theta y}^{2}|\beta_{\theta}\beta_{y}|^{2}}{(\gamma_{\theta}+\gamma_{y})^{2}+4\delta^{2}}$, we can get the evolution equation of the fluctuation of average phonon number of the translational mode,
\begin{eqnarray}
\frac{\rm{d}}{\rm{d}t}\langle\hat{n}_{y}\rangle=&-&\Big((1+\bar{n}_{y})\gamma_{y}+2\tilde{\eta}(1+\langle\hat{n}_{\theta}\rangle)\Big)\langle\hat{n}_{y}\rangle\nonumber\\
&+&\Big(\bar{n}_{y}\gamma_{y}+2\tilde{\eta}\langle\hat{n}_{\theta}\rangle\Big)(1+\langle\hat{n}_{y}\rangle)
\label{Eq:temporal evolution equation of ny}.
\end{eqnarray}
And for the librational mode, we can have the similar evolution equation of the fluctuation of average phonon number of the librational mode,
\begin{eqnarray}
\frac{\rm{d}}{\rm{d}t}\langle\hat{n}_{\theta}\rangle=&-&\Big((1+\bar{n}_{\theta})\gamma_{\theta}+2\tilde{\eta}(1+\langle\hat{n}_{y}\rangle)\Big)\langle\hat{n}_{\theta}\rangle\nonumber\\
&+&\Big(\bar{n}_{\theta}\gamma_{\theta}+2\tilde{\eta}\langle\hat{n}_{y}\rangle\Big)(1+\langle\hat{n}_{\theta}\rangle).
\label{Eq:temporal evolution equation of ntheta}
\end{eqnarray}

By solving Eq.(\ref{Eq:temporal evolution equation of ny}) and (\ref{Eq:temporal evolution equation of ntheta}), we can get the average fluctuations of the steady state phonon number for both translational and librational modes.
\begin{eqnarray}
\langle\hat{n}_{y}\rangle&=&\bar{n}_{y}-\frac{64\gamma_{\theta}(\gamma_{\theta}+\gamma_{y})\eta_{\theta y}^{2}\tilde{n_{\theta}}\tilde{n_{y}}}{\gamma_{\theta}\gamma_{y}\big((\gamma_{\theta}+\gamma_{y})^{2}+4\delta^{2}\big)+64(\gamma_{\theta}+\gamma_{y})^{2}\eta_{\theta y}^{2}\tilde{n}_{\theta}\tilde{n}_{y}}(\overline{n}_{y}-\overline{n}_{\theta}),\nonumber\\
\langle\hat{n}_{\theta}\rangle&=&\bar{n}_{\theta}+\frac{64\gamma_{y}(\gamma_{\theta}+\gamma_{y})\eta_{\theta y}^{2}\tilde{n_{\theta}}\tilde{n_{y}}}{\gamma_{\theta}\gamma_{y}\big((\gamma_{\theta}+\gamma_{y})^{2}+4\delta^{2}\big)+64(\gamma_{\theta}+\gamma_{y})^{2}\eta_{\theta y}^{2}\tilde{n}_{\theta}\tilde{n}_{y}}(\overline{n}_{y}-\overline{n}_{\theta}).
\label{Eq:average phonon number fluctuation of ntheta and ny}
\end{eqnarray}
For simplicity, we set $\langle\hat{n}_{y}\rangle=n_{y}$ and $\langle\hat{n}_{\theta}\rangle=n_{\theta}$. Because the translational mode and the librational mode are in the same temperature before cooling, the average excitation number $\bar{n}_\theta$ of librational mode is much less than  the translational mode average excitation number $\bar{n}_y$. Therefore, the translational mode can be cooled and the librational mode is heated after synthetic cooling. The cooling ratio $\xi=n_{y}'/\overline{n}_{y}$ could be used to
qualify the cooling.

\begin{figure}
\centering
\subfigure{
\begin{minipage}{0.45\textwidth}
\includegraphics[width=1\textwidth]{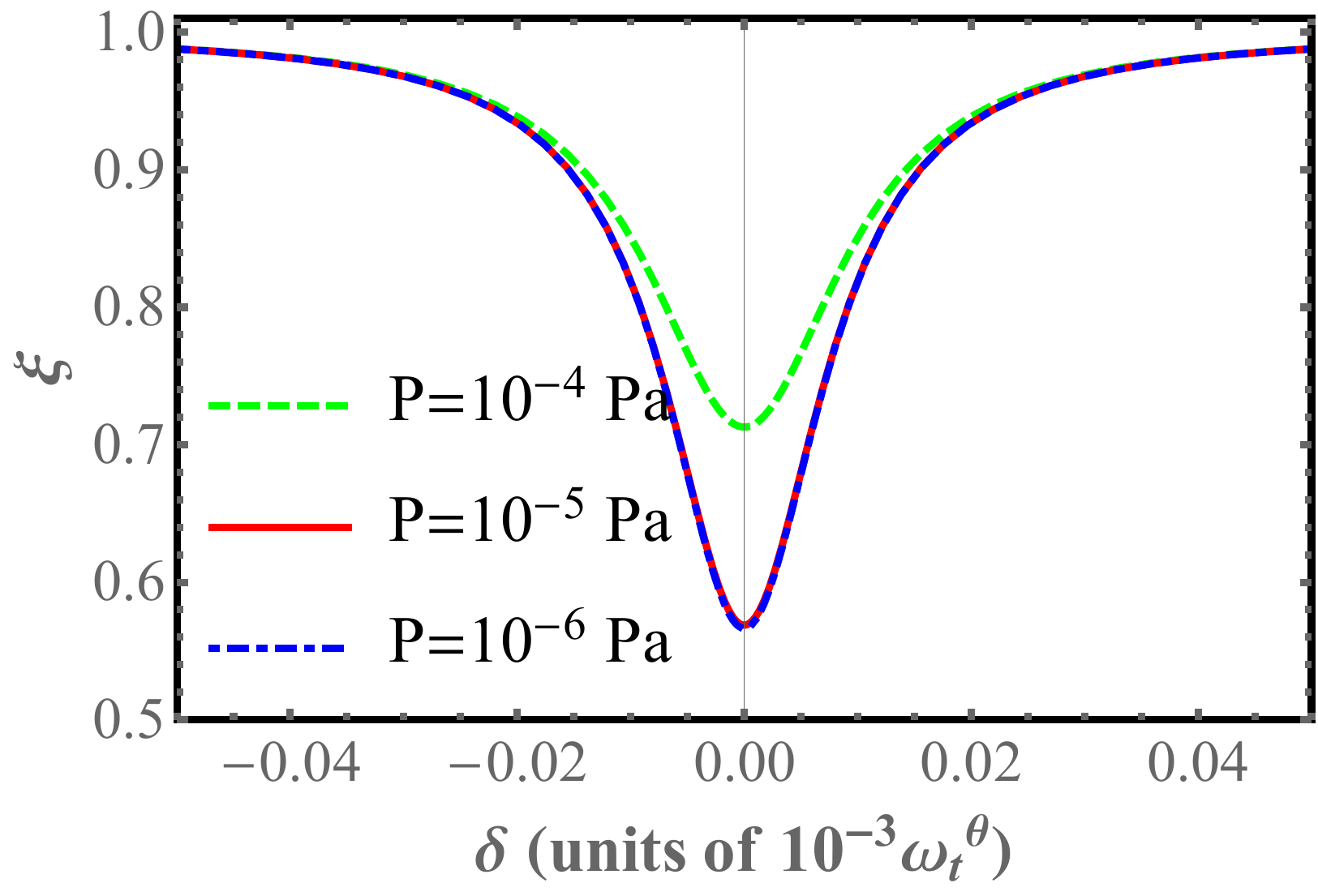}
\end{minipage}}
 \caption{Relation between the cooling ratio $\xi$ and $\delta$ for different residual air pressure. Different line present different pressure and environment temperature is $300~K$. The drive amplitude $\Omega_{1}=0.1$ and $\Omega_{2}=0.01$. $\Delta_{1}=10^{-3}$ and $\Delta_{2}=10^{-3}\omega_{t}^{y}$. These parameters are in units of $\omega_{t}^{\theta}$. The long axis and the short axis are respectively $50~nm$ and $25~nm$. The power and the waist of laser beam are $100~mW$ and $0.6~\mu m$ respectively. }
\label{fig:cooling phi different P}
\end{figure}

As previously mentioned, the translational mode is cooled and the cooling ratio ($\xi$) is determined by the decay of translational and librational mode, $\gamma_{y}$ and $\gamma_{\theta}$. These decays depend on both the residual air pressure ($P$) and the environment temperature $T$. The difference ($\delta$) between $\Delta_{{\rm eff}1}$ and $\Delta_{{\rm eff}2}$, is also important for cooling. The steady state phonon number of both the librational and translational mode ($\tilde{n}_{\theta}$ and $\tilde{n}_{y}$) are controlled by the driving amplitudes ($\Omega_{1}$ and $\Omega_{2}$). Therefore, we should consider the effect of the driving amplitudes for cooling.  At first, we fix the driving amplitude ($\Omega_{1}=0.1$ and $\Omega_{2}=0.1$) and the environment temperature ($T=300~K$). The cooling ratio $\xi$ is only determined by residual air pressure($P$) and $\delta$, as shown like Fig.~\ref{fig:cooling phi different P}.
The numerical solution presents that the higher vacuum is better for cooling. Nevertheless, when the residual air pressure is higher than $10^{-5}~Pa$, the cooling ratio is saturated. The optimal cooling takes place when the effective detuning $\Delta_{{\rm eff}1}$ and $\Delta_{{\rm eff}2}$ match each other perfectly.

\begin{figure}
\centering
\subfigure{
\begin{minipage}{0.45\textwidth}
\includegraphics[width=1\textwidth]{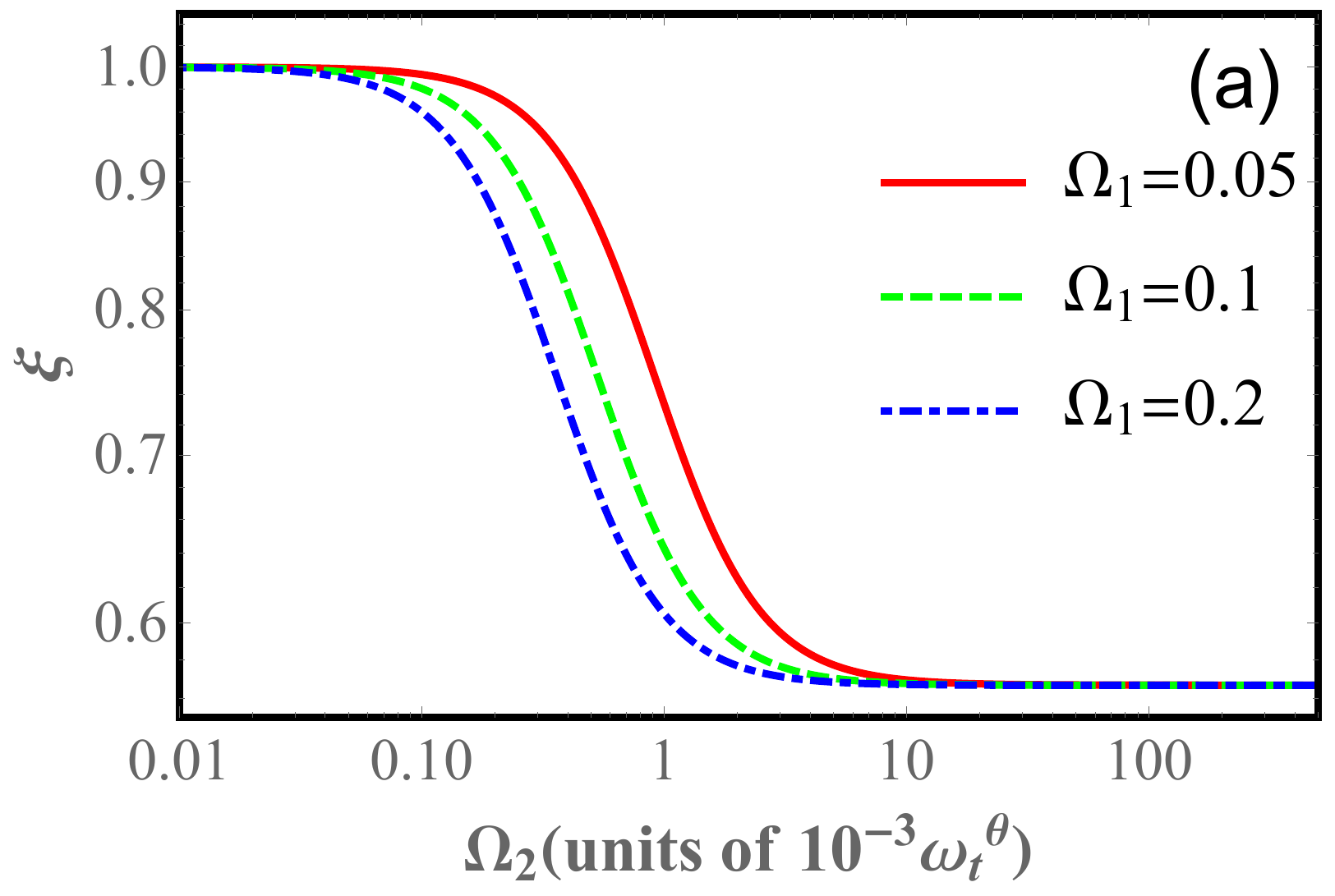}
\end{minipage}}
\subfigure{
\begin{minipage}{0.45\textwidth}
\includegraphics[width=1\textwidth]{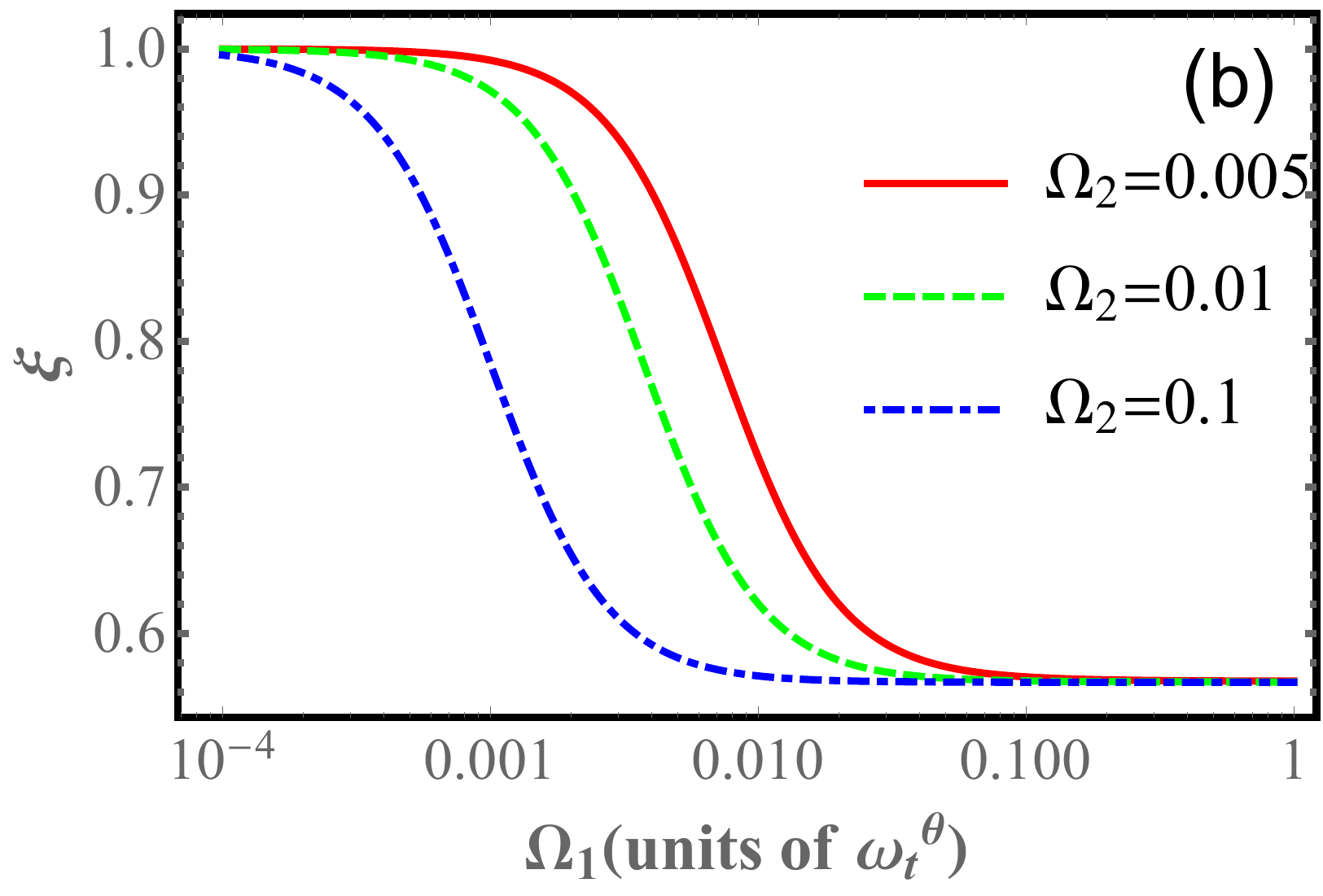}
\end{minipage}}
 \caption{The relatiob between the cooling ratio of the translatioanl mode and the drive amplitudes. $\Omega_{1}$ is fixed in (a)) and $\Omega_{2}$ is fixed in (b). The residual air pressure $P$ is $10~\mu \rm Pa$ and temperature $T$ is $300~K$. The size of particle and trapping laser is the same with Fig.~\ref{fig:cooling phi different P}. }
\label{fig:cooling Omega1 and Omega2}
\end{figure}

Driving amplitudes, $\Omega_{1}$ and $\Omega_{2}$, can affect the steady phonon number $\tilde{n}_{\theta}$ and $\tilde{n}_{y}$, the driving amplitudes also affect the cooling ratio $\xi$. Taking Fig.~\ref{fig:cooling Omega1 and Omega2} (a) for and example, when the driving amplitude of librational mode is fixed, the increment of $\Omega_{2}$ is good for cooling. $\Omega_{1}$ also plays part in this process. When $\Omega_{1}$ is small, although increment $\Omega_{2}$ is good for cooling, the cooling ratio will reach limit and trend to constant when $\Omega_{2}$ is larger than $0.01$, for example $\Omega_{1}=0.05$ in Fig.~\ref{fig:cooling Omega1 and Omega2}(a). Similar phenomenon also happens when $\Omega_{1}$ is changing for fixed $\Omega_{2}$, for example $\Omega_{2}=0.01$ in Fig.~\ref{fig:cooling Omega1 and Omega2}(b).
Therefore, the driving amplitude of these two mode should cooperate for cooling. When the driving amplitude increases, the cooling ratio will  gradually saturates shown as the dashed and doted dashed lines in Fig.~\ref{fig:cooling Omega1 and Omega2}. When $\Omega_{1}$ or $\Omega_{2}$ are fixed and optimized for cooling, the increment of $\Omega_{2}$ or $\Omega_{1}$ cannot remarkably strengthen cooling effect. Fig.~\ref{fig:cooling Omega1 and Omega2} shows that the cooling ratio is $\xi=0.57$.

In order to get higher cooling effect, the feedback cooling can be used. When the cooling ratio of the translational mode saturates, the feedback cooling scheme can improve the cooling ratio further. The fluctuation of steady state phonon number of translational mode after feedback cooling reads
\begin{eqnarray}
n_{y}'&=&\bar{n}_{y}-\frac{64(\gamma_{\theta}+\gamma_{\rm{fb}})(\gamma_{\theta}+\gamma_{\rm{fb}}+\gamma_{y})\eta_{\theta y}^{2}\tilde{n_{\theta}}'\tilde{n_{y}}'}{\gamma_{y}(\gamma_{\theta}+\gamma_{{\rm fb}})\big((\gamma_{\theta}+\gamma_{{\rm fb}}+\gamma_{y})^{2}+4\delta^{2}\big)+64(\gamma_{\theta}+\gamma_{{\rm fb}}+\gamma_{y})^{2}\eta_{\theta y}^{2}\tilde{n}_{\theta}\tilde{n}_{y}}(\overline{n}_{y}-\overline{n}_{\theta})
\label{Eq:average phonon number ny in feedback cooling}.
\end{eqnarray}
where $n_{y}'$ is the fluctuation of steady state phonon number of the translational mode under the feedback cooling and $\tilde{n}'_{\theta/y}$ is the steady state phonon number of librational mode or translational mode.
In this scheme, the residual air pressure $P$ and environment ($T$) are fixed, the decays of librational and translational mode do not change. By increasing the feedback strength, the synthetic cooling ratio can be promoted further for the fixed driving amplitudes. Meanwhile, because driving amplitudes directly affect the steady state average phonon numbers, $\tilde{n}'_{\theta}$ and $\tilde{n}'_{y}$, the synthetic cooling ratio is also determined by driving amplitudes of the translational and librational modes. For depicting more clearly, the decay caused by feedback $\gamma_{\rm fb}$ can be in units of $\gamma_{\theta}$.
\begin{figure}[htbp]
\centering
\subfigure{
\begin{minipage}{0.45\textwidth}
\includegraphics[width=1\textwidth]{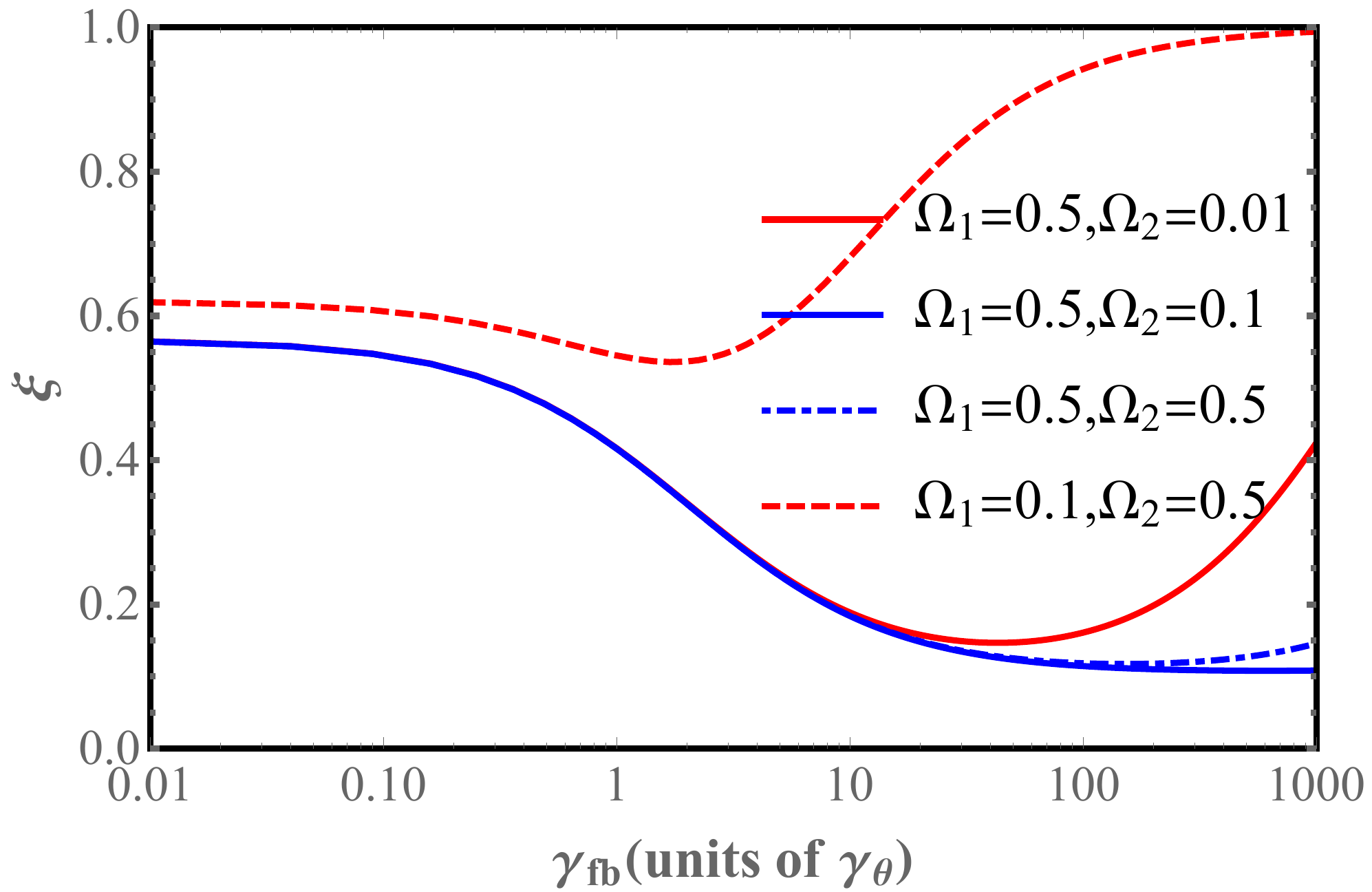}
\end{minipage}}
 \caption{Cooling ratio of transitional mode in different driving amplitude. Different line presents different drive amplitude of librational mode and translational mode. The parameters are the same to those in Fig.~\ref{fig:cooling phi different P}. }
\label{fig:cooling feedback}
\end{figure}

As shown in Fig.~\ref{fig:cooling Omega1 and Omega2}, the synthetic cooling of translational mode will saturate when the driving amplitudes increase. The feedback cooling scheme can break through the saturation of synthetic cooling and promote the cooling ratio in Fig.~\ref{fig:cooling feedback}. Because driving amplitudes is also an important aspect for the synthetic cooling of translational mode, $\Omega_{1}$ and $\Omega_{2}$ will affect the cooling ratio. When $\Omega_{1}$ and $\Omega_{2}$ do not match each other, the synthetic cooling ratio will not be significantly promoted and even the translational mode will be heated with the increment of feedback strength. For example, the red-dashed line in Fig.~\ref{fig:cooling feedback}. And when the driving amplitudes match each other, the synthetic cooling ratio can be significantly improved and even reaches one percent of the ambient temperature shown as blue line in Fig.~\ref{fig:cooling feedback}. In a word, feedback cooling can effectively improve the synthetic cooling ratio when driving amplitudes match each other or else feedback cooling will heat the translational mode.

%

\section{Discussion and Conclusion}
\label{SectV}

In the last section, the beam-splitter Hamiltonian was obtained by adjusting the driving frequency for the cooling of translational mode. And when $\Delta_{1}-24\eta_{\theta}|\beta_{\theta}|^{2}-4\eta_{\theta y}(|\beta_{y}|^{2}+1)=-\big(\Delta_{2}-24\eta_{y}|\beta_{y}|^{2}-4\eta_{\theta y}(|\beta_{\theta}|^{2}+1)\big)+\varphi$ with constant $\varphi$, the linearised Hamiltonian Eq.(\ref{Eq:linearisated Hamiltonian1}) can be transformed to the two-modes squeezing Hamiltonian as follow:
\begin{eqnarray}
\hat{H}_{\rm{tms}}&=&-4\hbar\eta_{\theta y}(\beta_{\theta}\beta_{y}\hat{b}_{\theta}^{\dag}\hat{b}_{y}^{\dag}e^{i\varphi}+{\rm H.C.})
\label{Eq:two mode Hamiltonian}.
\end{eqnarray}
In this way, the two-mode squeezing between the librational and the translational modes can also be generated, similarly as the Ref. ~\cite{Tan2013PRA,Pontin2016Prl,cai2017SCP}.

In this paper we systematically studied the coupling nonlinearity between librational mode and translational mode of an optically levitated ellipsoidal nanoparticle. The coupling of librational mode and translational mode is small, but it should not be neglected when proper driving is applied.  For coupling the Hamiltonian of these two motive modes, the stable-state analysis shows the driven librational mode and translational mode could have coupling bistability and one red-detuning drive of any mode could also stimulate the bistablity of other mode. In order to stabilize the system, the drives on the librational and the translational modes should be both blue-detuned. For the linearized coupling Hamiltonian between librational mode and translational mode, the synthetic cooling can be realised in steady state regime. To cool the translational mode by the librational mode, the lower pressure of air residual is always not helpful to translational mode cooling, and the cooling efficiency can be saturated when pressure decreases. The driving amplitude of these two modes is also important for sympathetic cooling, and the driving amplitude matching each other can increase the cooling efficiency of translational mode. However, here the cooling ratio is only $0.57$ of the initial temperature. To solve the problem we used the feedback cooling about librational mode for cooling the translational mode. This scheme breaks through the cooling ratio and improves the cooling efficient remarkably, and even reaches one percent of the ambient temperature. These investigations give us a new platform for the preparing macroscopic ground state, quantum information processing, etc.

\begin{acknowledgments} 
This work is supported by the NSFC grants (No.11374032, 61435007, 11534002), the Joint Fund of the Ministry of Education of China (6141A02011604), National Basic Research Program of China (Grant No. 2016YFA0301201), Science Challenge Project (No.TZ2018003) and NSAF (No.U1530401). We thank Prof. Tongcang Li for helpful discussions.
\end{acknowledgments}

\appendix
\section{Coupling bistability in Different detuning region}
\label{SectVI}

\begin{figure}[htbp]
\centering
\subfigure{
\begin{minipage}{0.45\textwidth}
\includegraphics[width=1\textwidth]{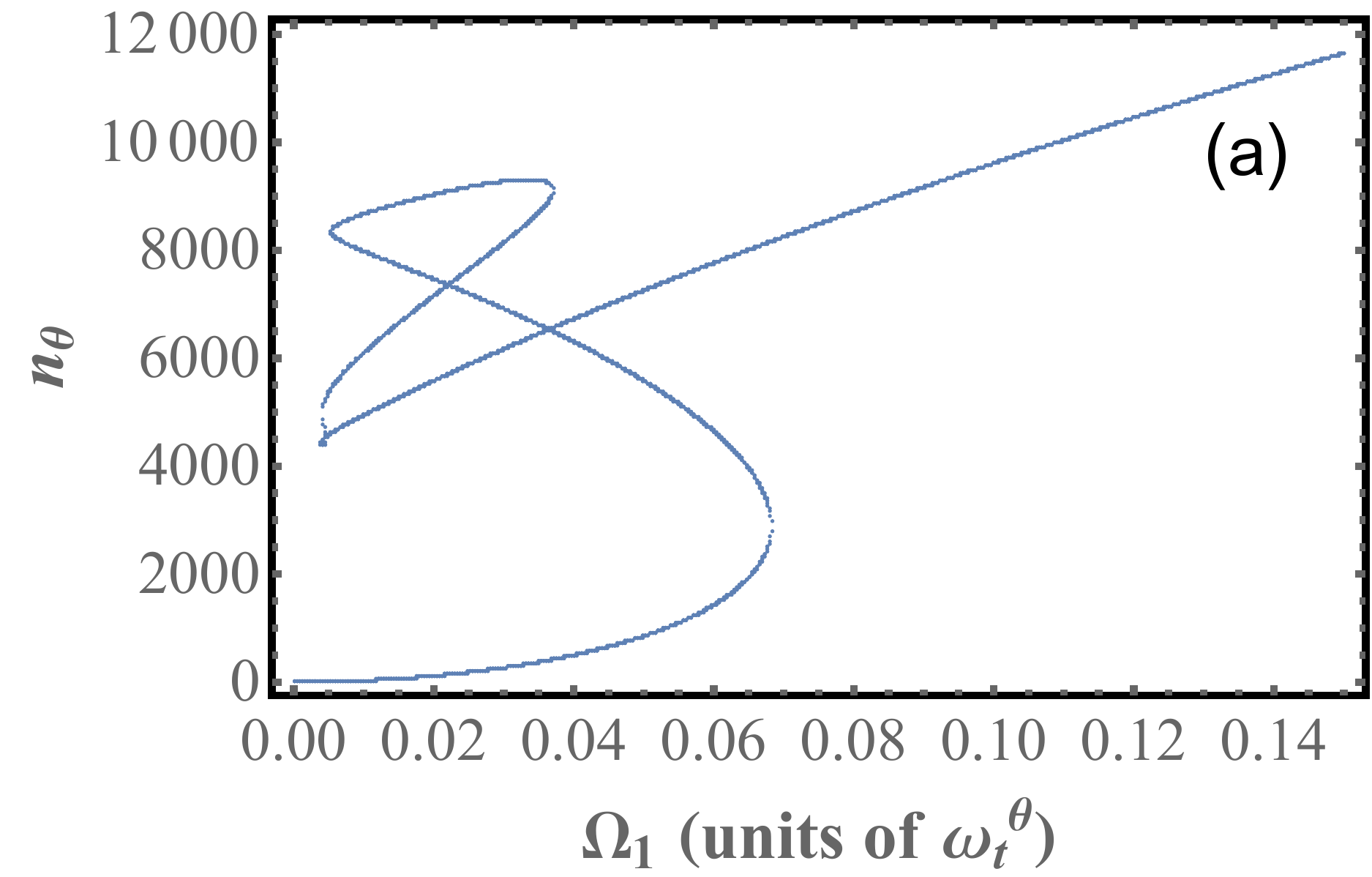}
\label{fig:multistabilityless1Omega1lib}
\end{minipage}}
\subfigure{
\begin{minipage}{0.45\textwidth}
\includegraphics[width=1\textwidth]{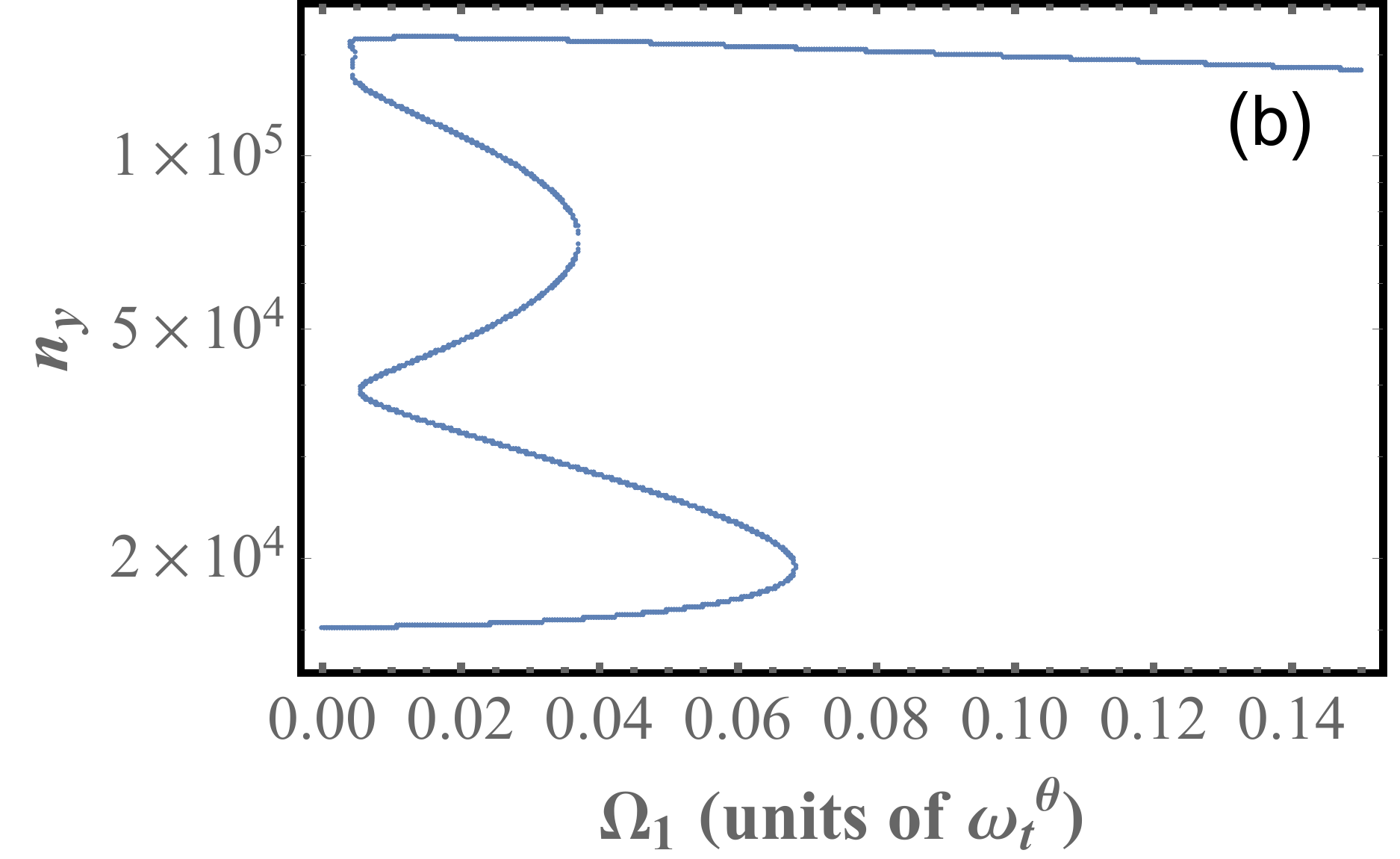}
\label{fig:multistabilityless1Omega1com}
\end{minipage}}
\caption{Multi-stability of $\tilde{n}_{\theta}$ and $\tilde{n}_{y}$ in this system when $\Omega_{2}=0.025$. (a) The multi-stability of $\tilde{n}_{\theta}$ versus the driven amplitude $\Omega_{1}$ under red detuning versus $\omega_{l1}$ and $\omega_{l2}$. (b) The multi-stability of $\tilde{n}_{y}$ versus the driven amplitude $\Omega_{1}$ under red detuning about $\omega_{l1}$ and $\omega_{l2}$. Other parameters are the same to those in Fig.\ref{fig:multistability in red detuning of two drive}.}
\label{fig:sectional figure of multistability about Omega1}
\end{figure}

In Sec. \ref{coupling_bistability}, the red-detuning drive induces the bistability and multi-stability of this coupling system of librational mode and translational mode. Fig.~\ref{fig:multistability in red detuning of two drive} gives us an example of the  multi-stability for librational mode and translational mode when the driving frequencies are red-detuned.
To show this property more clearly,  the multi-stability of $n_{\theta}$ and $n_{y}$ about $\Omega_{1}$ are shown as Fig.~\ref{fig:sectional figure of multistability about Omega1} when $\Omega_{2}=0.025$ and $\Omega_{1}\leq 0.15$.

\begin{figure}[htbp]
\centering
\subfigure{
\begin{minipage}{0.45\textwidth}
\includegraphics[width=1\textwidth]{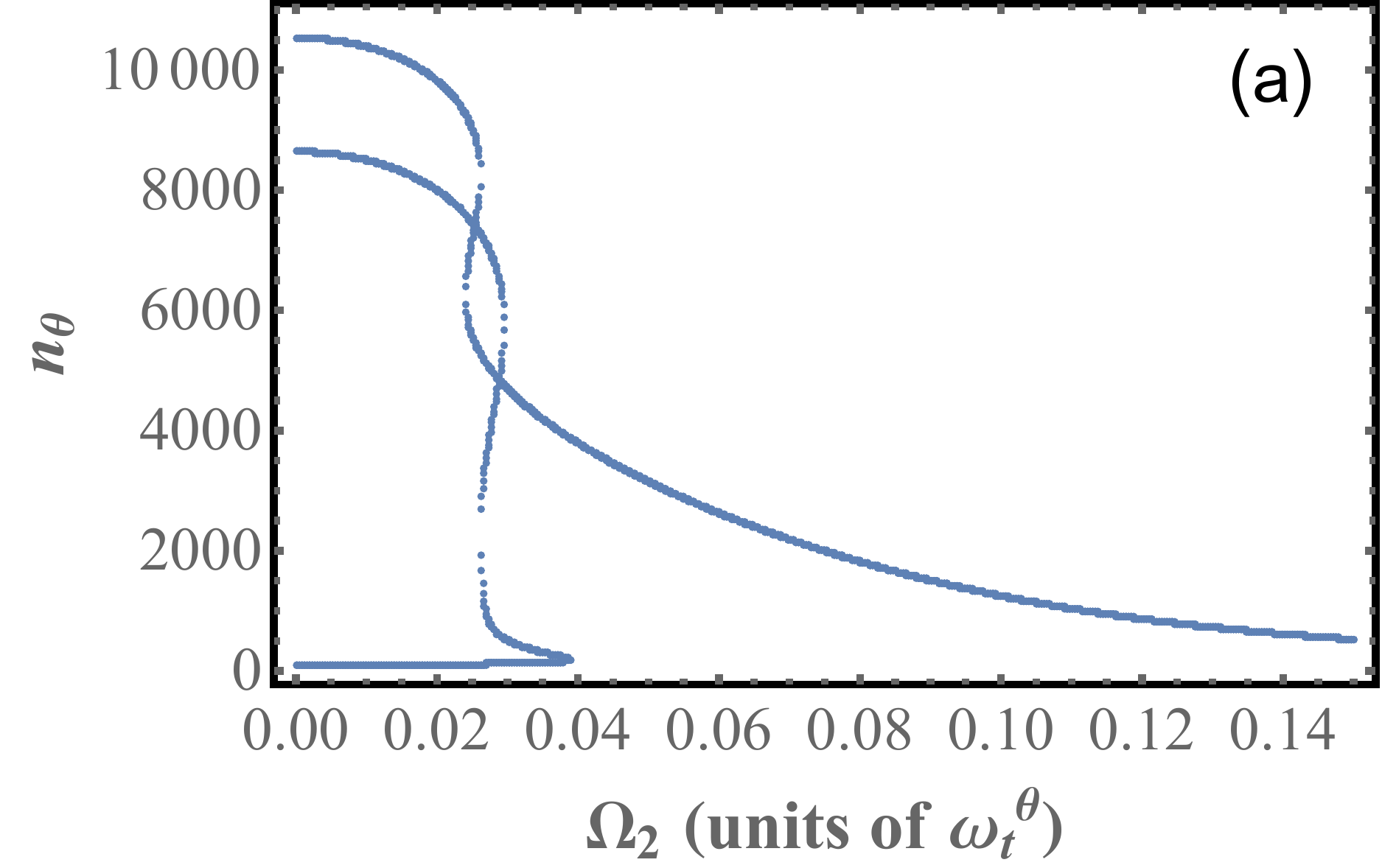}
\label{fig:multistabilityless1Omega2lib}
\end{minipage}}
\subfigure{
\begin{minipage}{0.45\textwidth}
\includegraphics[width=1\textwidth]{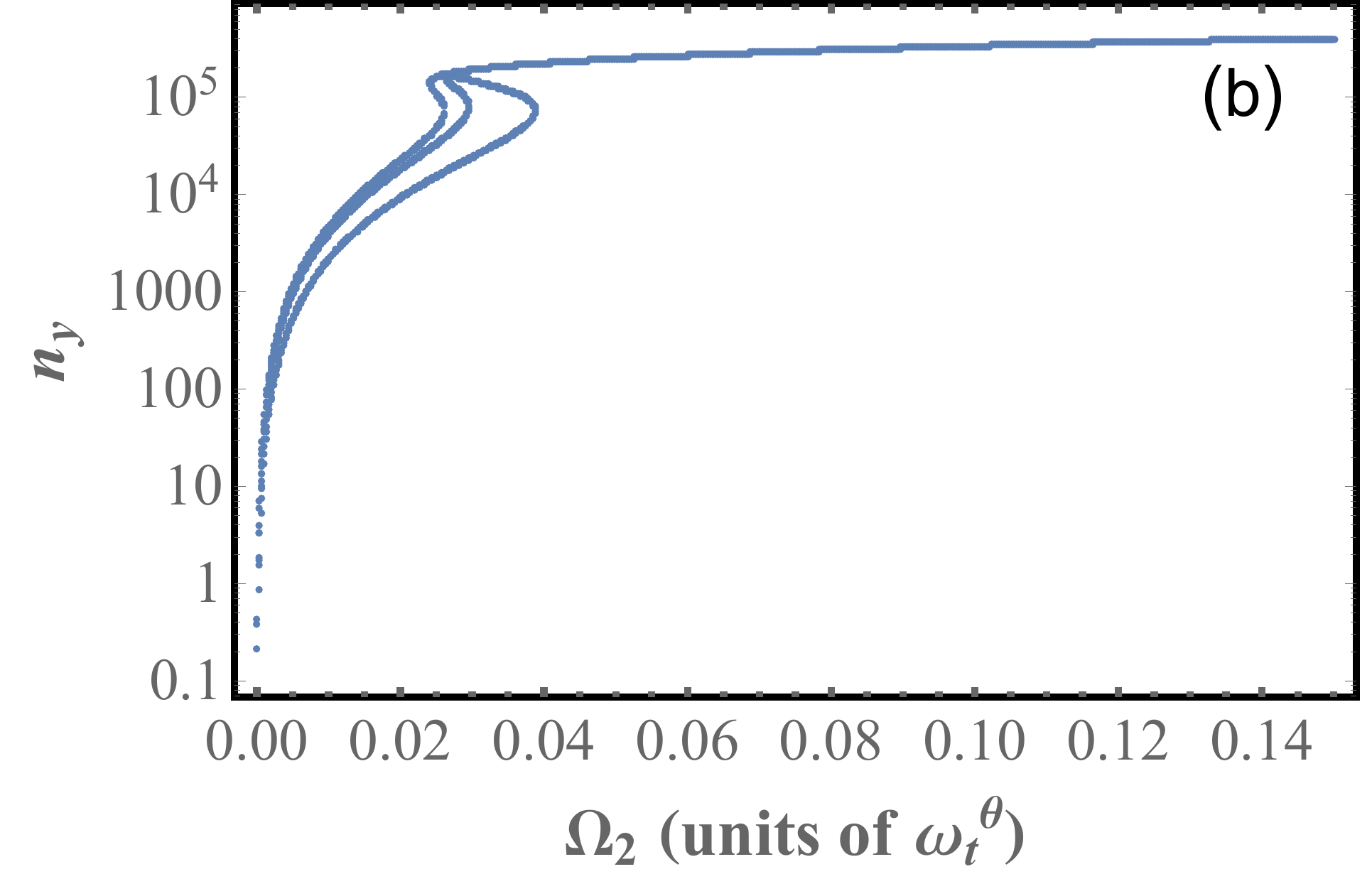}
\label{fig:multistabilityless1Omega2com}
\end{minipage}}
 \caption{Stability of $\tilde{n}_{\theta}$ and $\tilde{n}_{y}$ in this system when $\Omega_{1}=0.02$. (a) The stability of $\tilde{n}_{\theta}$ versus the driven amplitude $\Omega_{2}$ under red detuning about $\omega_{l1}$ and $\omega_{l2}$. (b) The stability of $\tilde{n}_{y}$ versus the driven amplitude $\Omega_{2}$ under red detuning about $\omega_{l1}$ and $\omega_{l2}$. Other parameters are the same to those in  Fig.\ref{fig:multistability in red detuning of two drive}.}
\label{fig:sectional figure of multistability about Omega2}
\end{figure}

At the same time, if we choose $\Omega_{1}=0.02$, the multi-stability of $\tilde{n}_{\theta}$ and $\tilde{n}_{y}$ about $\Omega_{2}$ is shown in Fig.~\ref{fig:sectional figure of multistability about Omega2}. When $\Omega_{2}>0.04$, $\tilde{n}_{\theta}$ is single valued (stable), and $\tilde{n}_{y}$ is also single valued (stable).

\begin{figure}[htbp]
\centering
\subfigure{
\begin{minipage}{0.45\textwidth}
\includegraphics[width=1\textwidth]{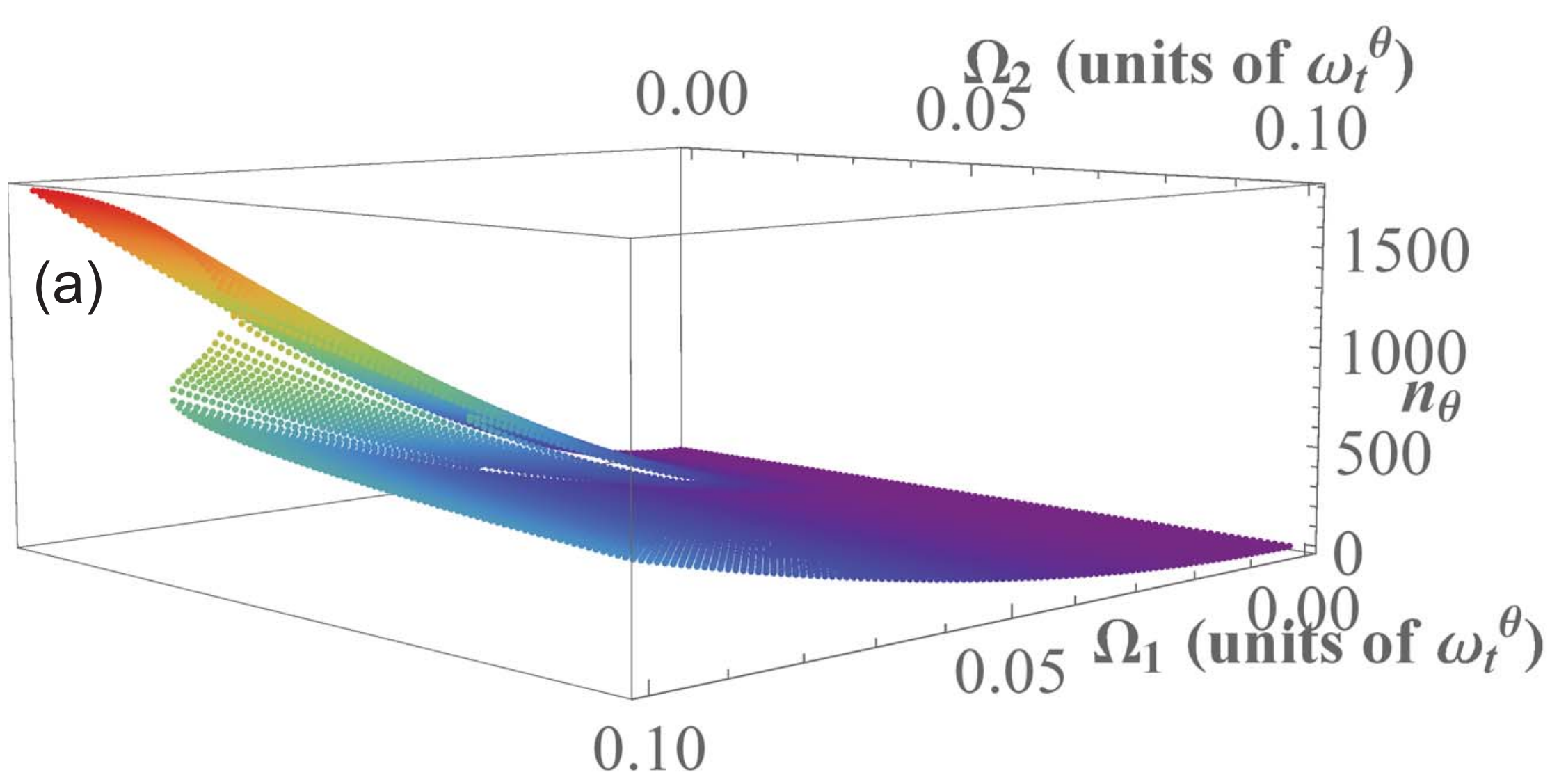}
\label{fig:multistabilityd1less0d2greater0lib}
\end{minipage}}
\subfigure{
\begin{minipage}{0.45\textwidth}
\includegraphics[width=1\textwidth]{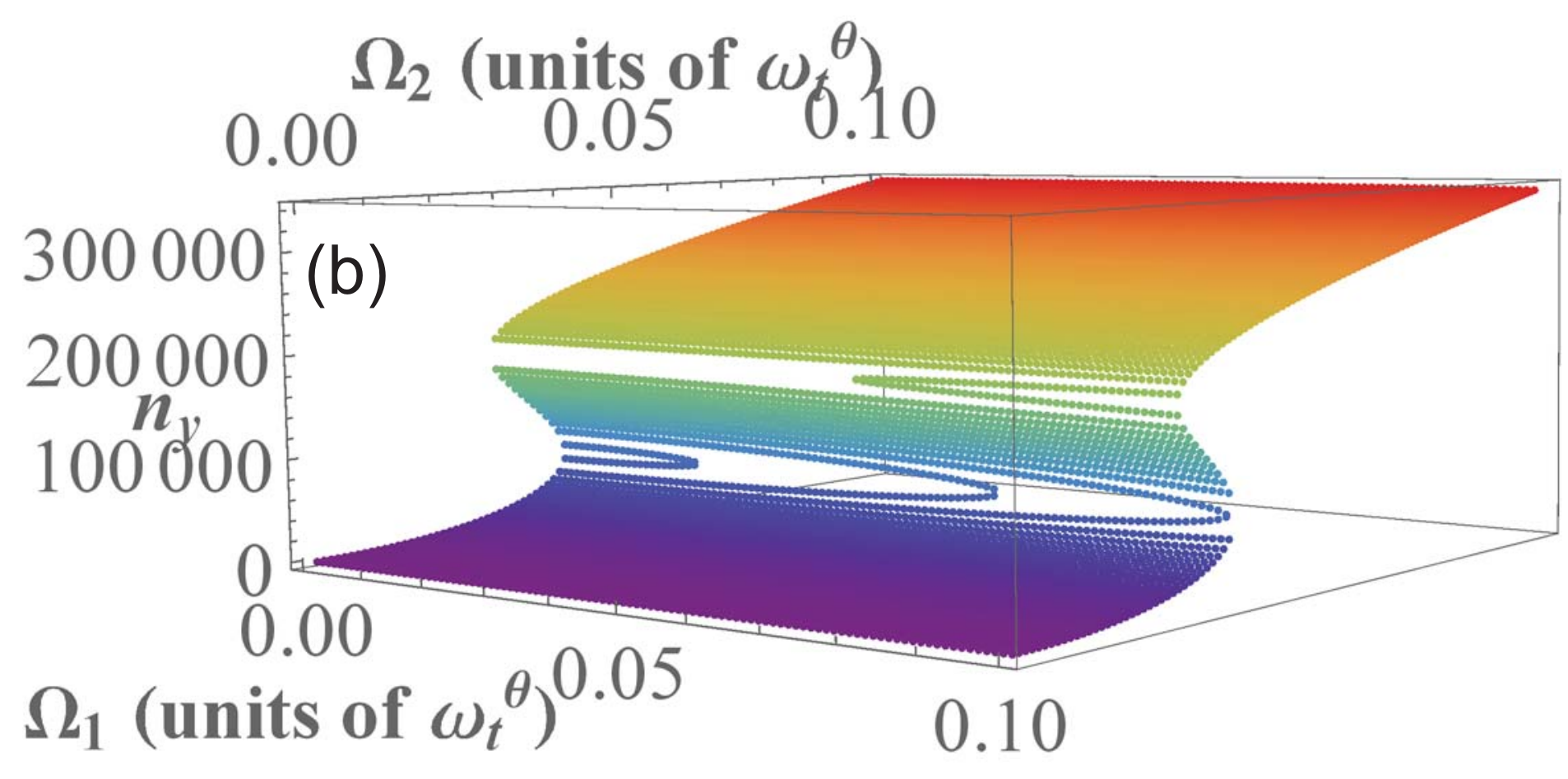}
\label{fig:multistabilityd1less0d2greater0com}
\end{minipage}}
 \caption{Bistability of librational mode and translational mode in $\Delta_{1}=-0.01$ and $\Delta_{2}=0.01\omega_{t}^{y}$ for this system. (a) the average phonon number of librational mode $\tilde{n}_{\theta}$ versus $\Omega_{1}$ and $\Omega_{2}$. (b) the average phonon number of translational mode $\tilde{n}_{y}$ versus $\Omega_{1}$ and $\Omega_{2}$. Other parameters are the same to those in Fig.\ref{fig:multistability in red detuning of two drive}.}
\label{fig:multistability d1less0 d2greater0}
\end{figure}

\begin{figure}[htbp]
\centering
\subfigure{
\begin{minipage}{0.41\textwidth}
\includegraphics[width=1\textwidth]{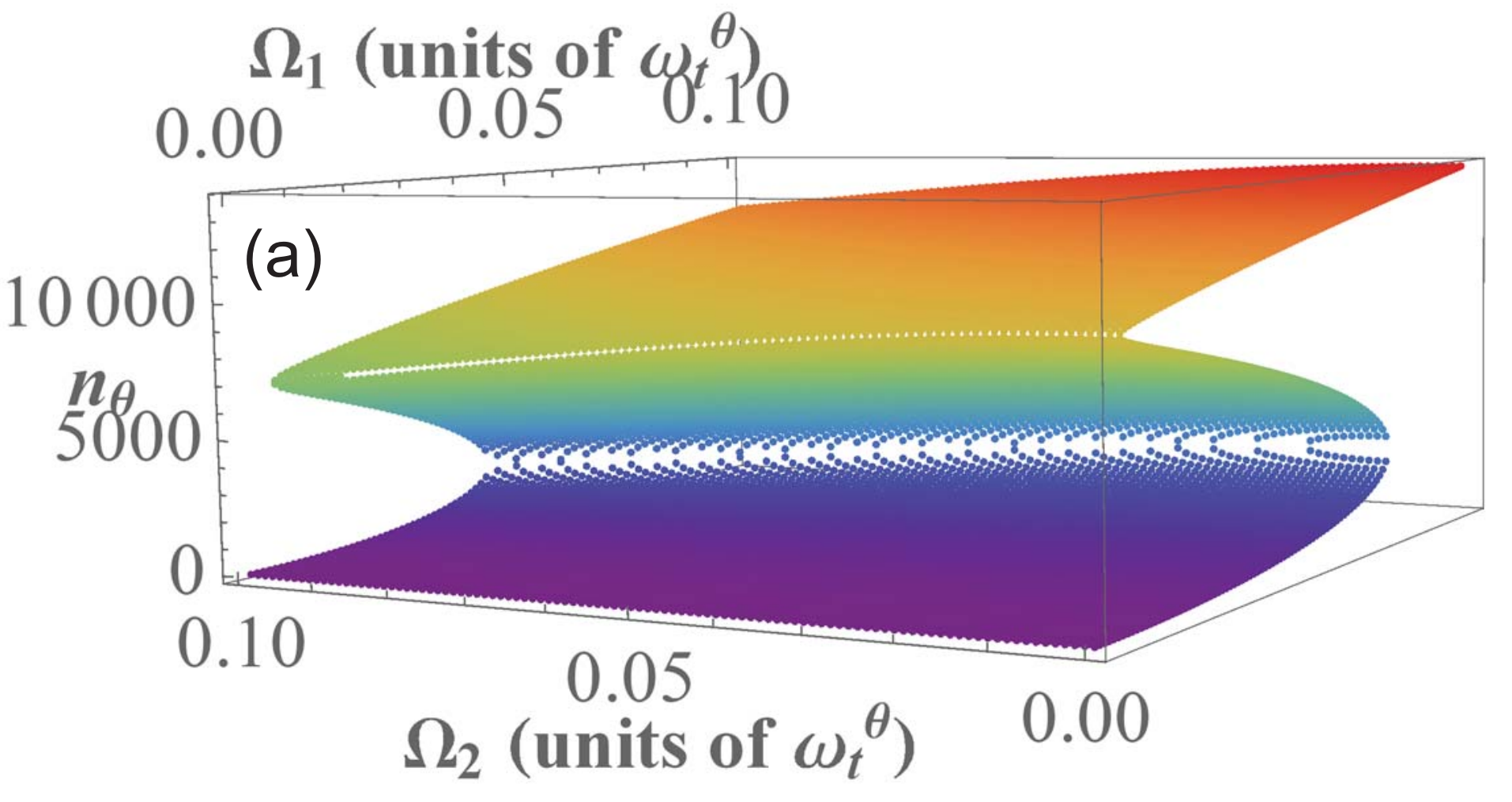}
\label{fig:multistabilityd1greater0d2less0lib}
\end{minipage}}
\subfigure{
\begin{minipage}{0.48\textwidth}
\includegraphics[width=1\textwidth]{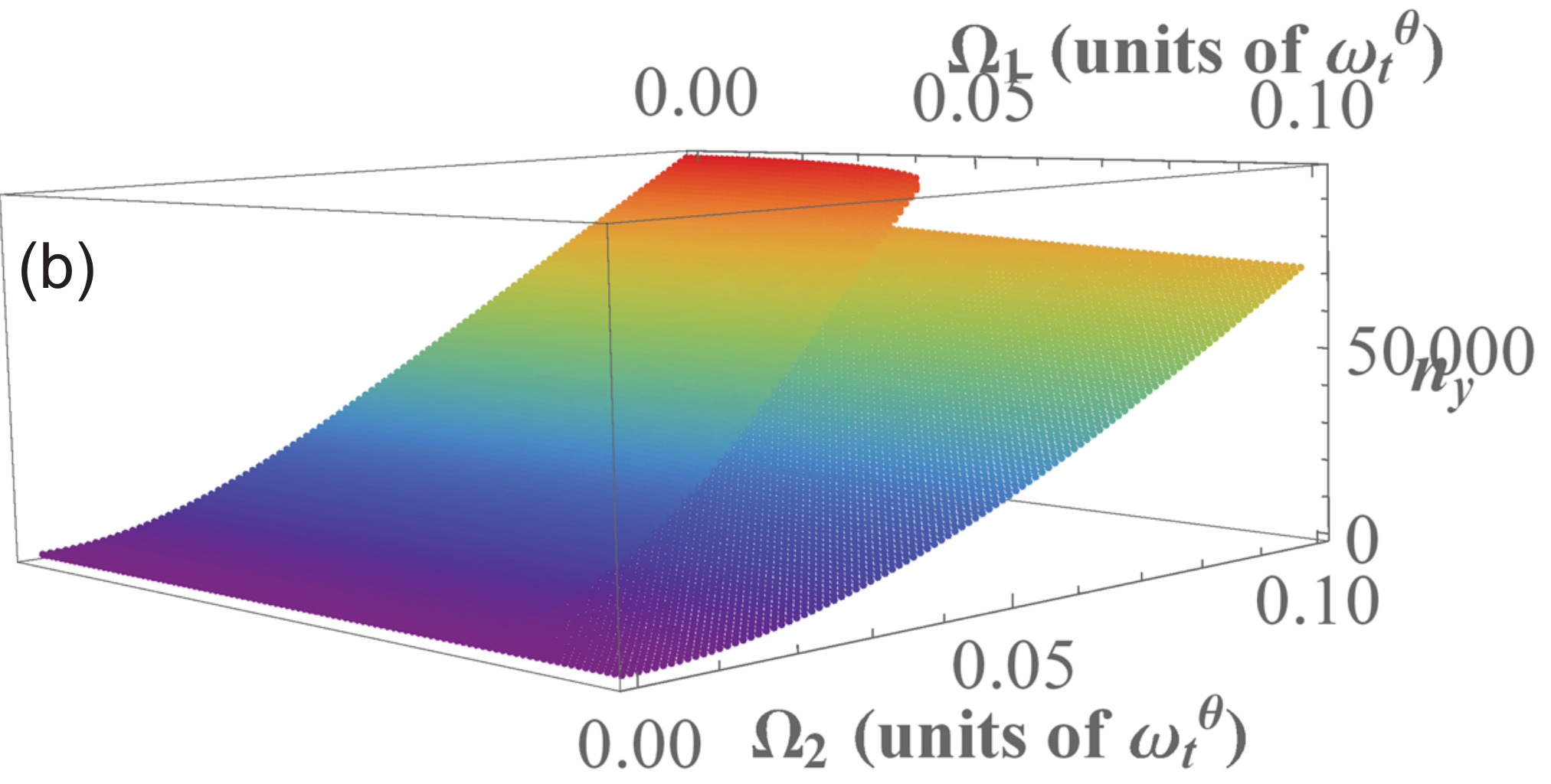}
\label{fig:multistabilityd1greater0d2less0com}
\end{minipage}}
 \caption{Bistability of librational mode and translational mode in $\Delta_{1}=0.01$ and $\Delta_{2}=-0.01\omega_{t}^{y}$ this system. (a) the average phonon number of librational mode $\tilde{n}_{\theta}$ versus $\Omega_{1}$ and $\Omega_{2}$. (b) the average phonon number of translational mode $\tilde{n}_{y}$ versus $\Omega_{1}$ and $\Omega_{2}$. Other parameters are the same to those in Fig.\ref{fig:multistability in red detuning of two drive}.}
\label{fig:multistability d1greater0 d2less0}
\end{figure}

When the effective driving detunings $\Delta_{1}\leq0$ and $\Delta_{2}\geq0$, $\tilde{n}_{\theta}$ and $\tilde{n}_{y}$ will have bistability, as shown in Fig.~\ref{fig:multistability d1less0 d2greater0}.
Similarly, when $\Delta_{1}\geq0$ and $\Delta_{2}\leq0$, $\tilde{n}_{\theta}$ and $\tilde{n}_{y}$ will present bistability as shown in Fig.~\ref{fig:multistability d1greater0 d2less0}.
However, in some parameter region of these two case, $\tilde{n}_{\theta}$ and $\tilde{n}_{y}$ have steady state property, for example, $\Omega_{2}\geq0.05$ in Fig.~\ref{fig:multistability d1less0 d2greater0} and $\Omega_{1}\geq0.08$ in Fig.~\ref{fig:multistability d1greater0 d2less0}, the average phonon number $\tilde{n}_{\theta}$ and $\tilde{n}_{y}$ have steady state property, it is useful for quantum measurement, quantum manipulation and so on.


\end{document}